\documentclass[a4paper,12pt]{article}
\usepackage{a4}
\usepackage{graphics}
\usepackage{epsfig}
\usepackage{amssymb}
\usepackage{color}

\newcommand{\la}{\lambda}

\newcommand{\eps}{\epsilon}

\newcommand{\chpm}{\tilde{\chi}^\pm}
\newcommand{\chmp}{\tilde{\chi}^\mp}

\newcommand{\chargino}{\tilde{\chi}^+}
\newcommand{\charginoplus}{\tilde{\chi}^+}
\newcommand{\charginominus}{\tilde{\chi}^-}

\newcommand{\chp}{\tilde{\chi}_1^+}
\newcommand{\chm}{\tilde{\chi}_1^-}

\newcommand{\chpj}{\tilde{\chi}_j^+}
\newcommand{\chmi}{\tilde{\chi}_i^-}

\newcommand{\neutralino}{\tilde{\chi}^0}

\newcommand{\sneutrino}{\tilde{\nu}}

\def\lsim{\raise0.3ex\hbox{$\;<$\kern-0.75em\raise-1.1ex\hbox{$\sim\;$}}}
\def\gsim{\raise0.3ex\hbox{$\;>$\kern-0.75em\raise-1.1ex\hbox{$\sim\;$}}}

\begin{document}
\hfill BONN-TH-2005-04\\
\hspace*{\fill} WUE-ITP-2005-010\\
\hspace*{\fill} hep-ph/0508267\\
\begin{center}
{\Large \bf 
	Higgs boson interference in $\mu^+\mu^-\to\charginoplus_i\charginominus_j$ 
	with longitudinally polarized beams
}			
\end{center}
\vspace{0.5cm}
\begin{center}
{\large \sc		
Olaf~Kittel$^a$, Federico~von~der~Pahlen$^b$}
\end{center}
\begin{center}
{\small \it  	$^a$ Physikalisches Institut der Universit\"at Bonn,
\\ 
		Nussallee 12, D-53115 Bonn, Germany
}
\end{center}
\begin{center}
{\small \it  	$^b$ Institut f\"ur Theoretische Physik und Astrophysik, Universit\"at W\"urzburg,
\\ 
		Am Hubland, D-97074 W\"urzburg, Germany
}
\end{center}
\begin{abstract}
We study chargino production at a muon collider with
longitudinally polarized beams and center of mass energies
around the heavy neutral Higgs boson resonances.
We show that the interference of the $CP$ even and $CP$ odd Higgs bosons 
can be analyzed using the energy distributions of the lepton or $W$ boson 
from the chargino two-body decays
$\tilde\chi^\pm_j\to\ell^\pm\tilde\nu_\ell$ or 
$\tilde\chi^\pm_j \to W^\pm \neutralino_1$, respectively.
The energy distributions depend on the longitudinal polarization of the 
decaying chargino, which are correlated to the muon beam polarizations. 
We define asymmetries in these energy distributions 
which allow a determination of 
the $H$ and $A$ couplings to the charginos 
and in particular of their relative phase.
We analyze the asymmetries, cross sections and branching ratios
in $CP$ conserving Minimal Supersymmetric Standard Model scenarios.
For  nearly degenerate Higgs bosons we find maximal 
asymmetries which can be measured with high statistical significance.


\end{abstract}
\newpage

\section{Introduction}

The $CP$ conserving Minimal Supersymmetric Standard Model (MSSM) 
contains three neutral Higgs bosons, a light scalar $h$, 
a heavier scalar $H$, and a pseudoscalar $A$~\cite{HK,GH,Djouadi}. 
A muon collider is an excellent tool to study the properties
of these neutral Higgs bosons, since they are resonantly produced in
s-channels~\cite{Djouadi,hefreports,mucolhiggs,barger1}.
A scan of the production line shape around the resonance region
allows the determination of e.g. the $H$ and $A$  masses and widths,
if the overlap of the two resonances is not too large~\cite{barger1}.
Moreover, if the polarizations of the muon beams \emph{and}  the final particles  
are taken into account, interference effects of the $H$ and $A$
channels give valuable information on the $CP$ properties of the Higgs 
bosons~\cite{asakawa}. The $H$--$A$ interference has been studied
recently in~\cite{Fraas:2004bq}, where the interactions of the Higgs 
bosons with neutralinos, the supersymmetric partners of the 
neutral Higgs and gauge bosons, have been analyzed.

In this paper we study the production of charginos,
the supersymmetric partners of the charged Higgs and gauge bosons, 
which allows precision measurements of the Higgs-chargino 
couplings~\cite{Fraas:2003cx}. 
We show that the longitudinal chargino polarizations  
are sensitive to the interference of the  $H$ and $A$ channels,
which is sizable if the two Higgs bosons are nearly degenerate,
i.e. if their mass difference is of the order of their decay widths. 
In order to probe the longitudinal chargino polarizations, we define
asymmetries in the energy distribution of the lepton $\ell$ from the
chargino decay 
$\tilde\chi_j^{\pm} \to \ell^{\pm}  \tilde\nu_\ell^{(\ast)}$, 
and the $W$ boson from the chargino decay
$	\tilde\chi_j^{\pm} \to W^{\pm}  \tilde\chi_k^0.$
A measurement of these asymmetries and the cross sections
allows a determination of the Higgs-chargino couplings.
In particular the asymmetries provide the relative phase 
between the $CP$ even and $CP$ odd Higgs boson couplings,
which would be a unique test of their $CP$ properties. 

In Section~\ref{Definitions and formalism} we give our 
definitions and formalism, and define the energy 
distribution asymmetries. 
In Section~\ref{Determination of the Higgs-chargino couplings} 
we study the dependence of these asymmetries on the Higgs-chargino 
couplings. 
In Section~\ref{Numerical results} we present numerical results and 
give a summary and conclusions in 
Section~\ref{Summary and conclusions}.


\section{Definitions and formalism
  \label{Definitions and formalism}}

We study pair production of charginos with momentum $p$ and helicity
$\lambda$
\begin{equation}
	\mu^+ + \mu^-\to \tilde\chi_i^{\mp}(p_{\chi_i^{\mp}},\lambda_i) 
						+ \tilde\chi_j^{\pm}(p_{\chi_j^{\pm}},\lambda_j)
	\label{production}
\end{equation}
with longitudinally polarized muon beams, 
and the subsequent leptonic two-body decay of one of the charginos
into a lepton and a sneutrino
\begin{equation}
	\tilde\chi_j^{\pm} \to \ell^{\pm} + \tilde\nu_\ell^{(\ast)}.
   \label{decay1}
\end{equation}
In the following we focus on the  case  $\ell = e,\mu$.
However, the results we obtain can be extended for $\ell = \tau$ 
and for the chargino decay into a $W$ boson and a neutralino,
\begin{equation}
	\tilde\chi_j^{\pm} \to W^{\pm} + \tilde\chi_k^0,
   \label{decay2}
\end{equation}
for which we give relevant formulas in Appendix~\ref{decdens}.

\subsection{Lagrangians and couplings}
%
\label{section:lagrdens}
The MSSM interaction Lagrangians for chargino production~(\ref{production}) 
via Higgs exchange are 
(in our notation we follow closely~\cite{HK,GH,Fraas:2004bq})
\begin{eqnarray}
\label{mumuphi}
{\cal L}_{\mu^+ \mu^- \phi} & = &
          g \, 
                \bar{\mu}\, 
          (c^{(\phi\mu)\,\ast} P_L + c^{(\phi\mu)} P_R)
		\, \mu \, \phi,\\
{\cal L}_{\tilde{\chi}^+  \tilde{\chi}^+ \phi} & = &
          g \, \bar{\tilde{\chi}}_i^+
          (c^{(\phi)}_{L\,ij} P_L + c^{(\phi)}_{R\,ij} P_R)
			 \tilde{\chi}_j^+\,\phi,
\label{chichiphi}
\end{eqnarray}
with $P_{R,L}=\frac{1}{2}(1\pm\gamma^5)$, $g$ the weak coupling constant 
and $\phi=H,A,h$.
The muon and chargino couplings to $H$ and $A$ are~\cite{GH}: 
\begin{eqnarray}
	c^{(H\mu)} &=& -\frac{m_\mu}{2 m_W}\frac{\cos\alpha}{\cos\beta},
\label{eq:cHmu}
\\
c^{(A\mu)} &=& i \frac{m_\mu}{2 m_W}\tan\beta,
\label{eq:cAmu}
\\
c^{(H)}_{L\,ij} &=&  - Q_{ij}^\ast\cos\alpha  -  S_{ij}^\ast\sin\alpha,
\label{eq:cH}
\\         
c^{(A)}_{L\,ij} &=& i(  Q_{ij}^\ast\sin\beta +  S_{ij}^\ast\cos\beta),
\label{eq:cA}
\\
c^{(\phi)}_{R\,ij} &=&c^{(\phi)\ast}_{L\,ji}, \quad \phi=H,A,
\label{eq:cLR}
\\
	Q_{ij} &=& \frac{1}{\sqrt{2}}U_{i2}V_{j1},
\\
	S_{ij} &=& \frac{1}{\sqrt{2}}U_{i1}V_{j2},
\end{eqnarray}
where $\alpha$ is the Higgs mixing angle, $\tan\beta=v_2/v_1$ is the ratio of 
the vacuum expectation values of the two neutral Higgs fields,
$\theta_W$ is the weak mixing angle and  $U$, $V$ are the $2\times 2$ 
matrices which diagonalize the chargino mass matrix $X$ with 
$U_{m \alpha}^* ~X_{\alpha\beta}~V_{\beta
	n}^{-1}= m_{\chi^\pm_m}~\delta_{mn}$~\cite{HK}.
The muon and chargino couplings to the lighter Higgs boson $h$ are obtained
by substituting $\alpha$ with $\alpha+\pi/2$ in~(\ref{eq:cHmu}) 
and~(\ref{eq:cH}).

The Lagrangian for chargino decay into a 
lepton and a sneutrino~(\ref{decay1}) is 
\begin{eqnarray}
	{\cal L}_{\ell \sneutrino_\ell \tilde{\chi}^+} & = & 
		- g V_{j1} \bar{\ell} P_R \tilde{\chi}_j^{+C} \sneutrino_\ell
	+	\mbox{h.c.},\qquad \ell=e,\mu.
\end{eqnarray}
The Lagrangians for $\ell=\tau$ and for the chargino decay 
into a $W$ boson and a neutralino~(\ref{decay2}) are given 
in Appendix~\ref{decdens}.

\subsection{Amplitudes and spin density matrix formalism}

For the calculation of the cross section for the combined process of 
chargino production~(\ref{production}) and decay, (\ref{decay1}) or (\ref{decay2}),  
which includes the chargino $\tilde\chi_j^\pm$ helicities $\lambda_j$,
we use the spin density matrix formalism of~\cite{Haber94}, 
as e.g. used for chargino production in $e^+ e^-$ annihilation 
in~\cite{gudi99}. The unnormalized spin density matrices 
$\rho^{P}$ of $\tilde\chi_i^{\mp}\tilde\chi_j^{\pm}$ production 
and $\rho^{D}$ of $\tilde\chi_j^{\pm}$ decay are given by
\begin{eqnarray}
	\rho^{P}_{\la_j\la^\prime_j} &=& \sum_{\la_i} 
		{T}^{P}_{\la_i\la_j}{{T}}^{P*}_{\la_i\la^\prime_j},
\label{rhop}
\\
	\rho^{D}_{\la^\prime_j\la_j} &=& 		
		{{T}}^{D\ast}_{\la^\prime_j}{{T}}^{D}_{\la_j},
\label{rhod}
\end{eqnarray} 
where $T^P_{\la_i\la_j}$ and $T^D_{\la_j}$ are the helicity amplitudes 
for production and decay, respectively. 
The amplitude squared for production and decay is then
\begin{eqnarray}
	|{{T}}|^2 &=& |\Delta(\tilde\chi_j^{\pm})|^2 \sum_{\la_j \la_j^\prime} 
	\rho^P_{\la_j \la_j^\prime}\rho^D_{\la_j^\prime\la_j},
\label{tsquare}
\end{eqnarray}
with the propagator
$\Delta(\tilde\chi_j^{\pm})=i/[p^2_{\chi_j^{\pm}}-m_{\chi_j^{\pm}}^2+
	i m_{\chi_j^{\pm}}\Gamma_{\chi_j^{\pm}}]$,
where $p_{\chi_j^{\pm}}$, $m_{\chi_j^{\pm}}$ and
$\Gamma_{\chi_j^{\pm}}$ denote the four-momentum, mass and
width of the chargino, respectively.

Introducing a set of chargino spin vectors $s_{\chi_j^{\pm}}^a$, 
given in Appendix~\ref{spinvectors},
the spin density matrices~(\ref{rhop}) and~(\ref{rhod})
can be expanded in terms of the Pauli matrices $\tau^a$
\begin{eqnarray}
	\rho^P_{\la_j \la_j^\prime} &=& 
		\delta_{\la_j \la_j^{\prime}}P + \sum_{a=1}^3 \tau_{\la_j \la_j^{\prime}}^a  {\Sigma}_P^{a},
\label{rhoP}
\\
	\rho^D_{\la_j^\prime \la_j} &=& 
		\delta_{\la_j^\prime \la_j}D + \sum_{a=1}^3 \tau_{\la_j^{\prime} \la_j}^a  {\Sigma}_D^{a}.
\label{rhoD}
\end{eqnarray}
With our choice of the spin vectors, ${\Sigma}_P^{3}/P$ is the 
longitudinal polarization of $\chpm_j$, 
${\Sigma}_P^{1}/P$ is the transverse polarization 
in the production plane and ${\Sigma}_P^{2}/P$ is the polarization
perpendicular to the production plane.
Inserting the density matrices~(\ref{rhoP}) and~(\ref{rhoD}) 
into~(\ref{tsquare}) gives
\begin{eqnarray}
	|{{T}}|^2 &=&  2|\Delta(\tilde\chi_j^{\pm})|^2 
	(P D + \sum_{a=1}^3 {\Sigma}_P^{a} {\Sigma}_D^{a}).
\label{eq:tsquare}
\end{eqnarray}
The first term in~(\ref{eq:tsquare}) is independent of the 
chargino polarization whereas the second term describes 
the spin correlations between production and decay. 
Cross sections and distributions are now obtained by integrating 
$|T|^2$ over the Lorentz invariant phase space 
element $d{\rm Lips}$ 
\begin{equation}
	d\sigma=\frac{1}{2 s}|T|^2d{\rm Lips},
\label{crossection}
\end{equation}
where we use the narrow width approximation for the 
propagator of the decaying chargino.
Explicit formulas of the phase space for chargino 
production~(\ref{production}) and decay, (\ref{decay1}) or (\ref{decay2}),
can be found e.g. in~\cite{Kittel:2005rp}. 

\subsubsection{Contributions from $H$ and $A$ exchange}

The expansion coefficients of the chargino production 
matrix~(\ref{rhoP}) subdivide into contributions from the 
Higgs resonances and the continuum, respectively, 
\begin{equation}
	P=P_r+P_{cont}, \qquad {\Sigma}_P^{a} ={\Sigma}_r^{a}
	+{\Sigma}_{cont}^{a}~.
\label{contributions}
\end{equation}
The continuum contributions $P_{cont}$, ${\Sigma}_{cont}^{a}$ are 
those from the non-resonant $\gamma$, $Z$ and $\tilde\nu_\mu$ 
exchange channels and can be found in~\cite{gudi99}.
The resonant contributions are those
from s-channel exchange of the Higgs bosons $H$ and $A$ 
\begin{eqnarray}
\label{proddens.cpc}
	P_{r} &=&
		\sum_{\phi=H,A}   P_{r}^{(\phi\phi)} +   P_{r}^{(HA)},
\\
	\Sigma_{r}^3 & = & \sum_{\phi=H,A}  \Sigma_{r}^{3\, (\phi\phi)} + 
	                                    \Sigma_{r}^{3\, (HA)} ,
 \end{eqnarray}
which read for $\mu^+\mu^-\to \tilde\chi_i^-\tilde\chi_j^+$
\begin{eqnarray}
	P_{r}^{(\phi\phi)}  &=& 
	\frac{g^4}{4} (1 + \mathcal{P}_+  \mathcal{P}_-)
			|\Delta(\phi)|^2 |c^{(\phi\mu)}|^2 
\nonumber
\\[1mm]& & 
	\left[
		(|c_{L}^{(\phi)}|^2+|c_{R}^{(\phi)}|^2) 
	 (s-m_{\chi_i^{\pm}}^2-m_{\chi_j^{\pm}}^2)
		-4 \mbox{Re}\{{c_{L}^{(\phi)} c_{R}^{(\phi)\,*}}\} 
		m_{\chi_i^{\pm}} m_{\chi_j^{\pm}}
	\right] s,
\quad
\label{pcpconsphiphi}
\\[1mm] 
  P_{r}^{(HA)} &=&  - 
		\frac{g^4}{2} (\mathcal{P}_+ + \mathcal{P}_-) 
		\mbox{Re}\{\Delta(H)\Delta(A)^*\} 
		\mbox{Im}\{c^{(H\mu)}c^{(A\mu)^*}\} 
\nonumber
\\[1mm]
& & 
	\Big[
	\mbox{Im}\{{c_{L}^{(H)} c_{L}^{(A)\,*}}+ {c_{R}^{(H)}c_{R}^{(A)\,*}}\}
	(s-m_{\chi_i^{\pm}}^2-m_{\chi_j^{\pm}}^2)
\nonumber\\ & &	
- 2\,\mbox{Im}\{{c_{L}^{(H)} c_{R}^{(A)\,*}}+{c_{R}^{(H)} c_{L}^{(A)\,*}}\}
	m_{\chi_i^{\pm}} m_{\chi_j^{\pm}} 
	\Big]
	s,
\label{pcpconsHA}
\\[2mm]
	\Sigma_{r}^{3\, (\phi\phi)}  & = & 
	\frac{g^4}{4} (1 + \mathcal{P}_+ \mathcal{P}_-) 
	|\Delta(\phi)|^2 |c^{(\phi\mu)}|^2 
	(|c_{L}^{(\phi)}|^2 - |c_{R}^{(\phi)}|^2) 
	s\sqrt{\lambda_{ij}},
\label{sigmacpconsphiphi}
\\[1mm]
	\Sigma_{r}^{3\, (HA)}  & = & 
	 -\frac{g^4}{2}\mbox{Re}\{\Delta(H)\Delta(A)^\ast\} 
	(\mathcal{P}_+ + \mathcal{P}_-) 
	\nonumber\\ 
		& & 
		\mbox{Im}\{
			c_{L}^{(H)} c_{L}^{(A)\ast} 
		-	c_{R}^{(H)} c_{R}^{(A)\ast}
		\}
	\mbox{Im}\{ c^{(H\mu)} c^{(A\mu)\ast}\}
	s\sqrt{\lambda_{ij}}.
\ \ 
\label{sigmacpconsHA}
\end{eqnarray}
The resonant contributions 
$\Sigma_{r}^1$ and $\Sigma_{r}^2$
to the transverse polarizations of the chargino vanish, 
since the s-channel exchange is due to scalar Higgs bosons.
In the above formulas the chargino indices of the couplings 
$c^{(\phi)}_{R} \equiv c^{(\phi)}_{R\,ij}$ and 
$c^{(\phi)}_{L}\equiv c^{(\phi)}_{L\,ij}$
have been suppressed, 
the longitudinal beam polarizations are denoted by 
$\mathcal{P}_+$, $\mathcal{P}_-$, and
\begin{eqnarray}
	\Delta(\phi) &=& i [(s-m_\phi^2) + i m_\phi \Gamma_\phi]^{-1},\qquad \phi=H,A,
\label{hpropagators}\\
	\lambda_{ij} 	&=&
	\lambda(s,m_{\chi_i^{\pm}}^2,m_{\chi_j^{\pm}}^2), 
\end{eqnarray}
with $\lambda(x,y,z) = x^2+y^2+z^2-2(xy+xz+yz)$.
Note that both $P_{r}^{(HA)}$ and $\Sigma_{r}^{3\, (\phi\phi)} $ 
vanish for production of equal charginos $i=j$
since  then the  Higgs-chargino couplings are parity conserving, 
with $c_{L\,ii}^{(\phi)} = c_{R\,ii}^{(\phi)\ast}$. 
These two terms are only present for $\chpm_1 \chmp_2$ production
since
$c_{L\,ij}^{(\phi)} \neq c_{R\,ij}^{(\phi)\ast}$ for $i\neq j$
in general.
We neglect interferences of the chirality violating Higgs 
exchange amplitudes with the chirality conserving continuum amplitudes, 
which are of order 
$m_\mu/\sqrt{s}$.
Further we neglect 
contributions from $h$ exchange 
far from its resonance.

	In order to find observables which are sensitive to
	the $H$--$A$ interference, we analyze the properties of the coefficients 
	$P$ and $\Sigma_P^3$~(\ref{contributions}) under 
	parity and charge conjugation.
For the production of the charge conjugated pair of charginos
$\mu^+\mu^-\to \tilde\chi_i^+\tilde\chi_j^-$
they transform into
\begin{eqnarray}
\Sigma_{cont}^3 &\to& - \Sigma_{cont}^3,
\label{Ccontdependence}
\\
P_{r}^{(HA)} &\to& - P_{r}^{(HA)}, 
\\
\Sigma_{r}^{3\, (\phi\phi)} &\to& - \Sigma_{r}^{3\, (\phi\phi)},
\label{Cresdependence}
\end{eqnarray}
while $P_{cont}$~\cite{gudi99}, $P_{r}^{(\phi\phi)}$ and 
$\Sigma_{r}^{3\, (HA)}$ do not change.
For equal beam polarizations 
$\mathcal{P}_+=\mathcal{P}_-\equiv \mathcal{P}$
the resonant contributions transform
under $\mathcal{P} \to - \mathcal{P}$ into
\begin{eqnarray}
P_r^{(HA)} &\to& - P_r^{(HA)}, 
\label{PPrdependence}
\\
\Sigma_{r}^{3\, (HA)} &\to& - \Sigma_{r}^{3\, (HA)},
\label{PSigmaHArdependence}
\end{eqnarray}
while the terms $P_r^{(\phi\phi)}$, $\Sigma_{r}^{3\, (\phi\phi)}$
and the continuum contributions $P_{cont}$ and $\Sigma^3_{cont}$~\cite{gudi99}, 
are invariant. 
	Note that the $H$--$A$ interference terms $P_r^{(HA)}$~(\ref{pcpconsHA})
	and $\Sigma_{r}^{3\, (HA)}$~ (\ref{sigmacpconsHA}) are parity odd and 
	thus vanish for zero beam polarizations $\mathcal{P}_+ = \mathcal{P}_-=0$.

\subsubsection{Chargino decay into electrons and muons}

The expansion coefficients of the chargino decay matrix~(\ref{rhoD}) 
for the chargino decay  $\tilde\chi_j^{+} \to \ell^{+} ~ \tilde\nu_\ell$, 
with  $\ell = e,\mu$, are
   \begin{eqnarray}
		D & = & \frac{g^2}{2} |V_{j1}|^2 
		(m_{\chi_j^{\pm}}^2 -m_{\tilde{\nu}_\ell}^2 ),
   \label{DR}   \\
      \Sigma^{a}_{D} &=&    
		 -g^2 |V_{j1}|^2 m_{\chi_j^{\pm}} (s^a_{\chi_j^{\pm}}
			\cdot p_{\ell}).
   \label{SigmaD}
\end{eqnarray}
The coefficient $\Sigma^{a}_{D}$ for the charge conjugated process, 
$\tilde\chi_j^- \to \ell^- ~ \tilde\nu_\ell^\ast$, 
is obtained by inverting the sign of~(\ref{SigmaD}).
The coefficients for $\ell = \tau$ and for chargino decay
into a $W$ boson and a neutralino are given in Appendix~\ref{decdens}.

\subsection{Energy distribution}

In the center of mass system (CMS), the kinematical limits of the energy 
of the decay particle $\lambda=e, \mu, \tau,W$ 
from the chargino decays~(\ref{decay1}) and~(\ref{decay2}) are 
\begin{eqnarray}
E_\lambda^{max(min)}&=& 	 \bar{E}_\lambda \pm \Delta_\lambda,
\label{kinlimits}
\end{eqnarray}
which read for the leptonic ($\lambda=\ell$) chargino decays 
\begin{eqnarray}
	\bar{E}_\ell &=& \frac{E_\ell^{max}+E_\ell^{min}}{2} = 
	\frac{ m_{\chi^{\pm}_j}^2-m_{\sneutrino_\ell}^2}{2 m_{\chi^{\pm}_j}^2} E_{\chi^{\pm}_j},
\label{ehalf}
\\
\Delta_\ell &=& \frac{E_\ell^{max}-E_\ell^{min}}{2} = 
\frac{ m_{\chi^{\pm}_j}^2-m_{\sneutrino_\ell}^2}{2 m_{\chi^{\pm}_j}^2}
|\vec{p}_{\chi^{\pm}_j}|,
\qquad \ell=e,\mu,\tau.
\label{edif}
\end{eqnarray}
With these definitions we can rewrite the factor 
$\Sigma^{3}_{D}$~(\ref{SigmaD}), that multiplies the 
longitudinal chargino polarization $\Sigma^{3}_{P}$ in~(\ref{eq:tsquare}), 
\begin{eqnarray}
	\Sigma^{3}_{D} &=& \eta_{\lambda\pm} \frac{D}{\Delta_\lambda}
	(E_\lambda-\bar{E}_\lambda),\quad \lambda=e,\mu,\tau,W,
\label{etal}
 \end{eqnarray}
where we have used 
\begin{eqnarray}
	m_{\chi_j^{\pm}} (s^3_{\chi_j^{\pm}}\cdot p_\lambda)=
	-\frac{m_{\chi_j^{\pm}}^2}{|\vec{p}_{\chi^{\pm}_j} |}
	(E_\lambda-\bar{E}_\lambda).
\end{eqnarray}
The factor $\eta_{\lambda^{\pm}}$ is a measure of 
parity violation, which is maximal $\eta_{\ell^{\pm}}=\pm1$  
for the decay 
$\tilde\chi_j^{\pm} \to \ell^{\pm} ~\tilde\nu_\ell^{(\ast)}$, 
for $\ell=e,\mu$, since the sneutrino couples purely left handed.
For $\ell=\tau$ or for chargino decays into a $W$ and a neutralino, 
the factors $\eta_{\tau^{\pm}}$~(\ref{etatau}) and 
$\eta_{W^{\pm}}$~(\ref{etaW}), respectively,  are generally smaller, 
thus reducing $\Sigma^{3}_{D}$.

\begin{figure}[t]
\centering
\begin{picture}(14,4.6)
\put(.7,-11.95){\includegraphics{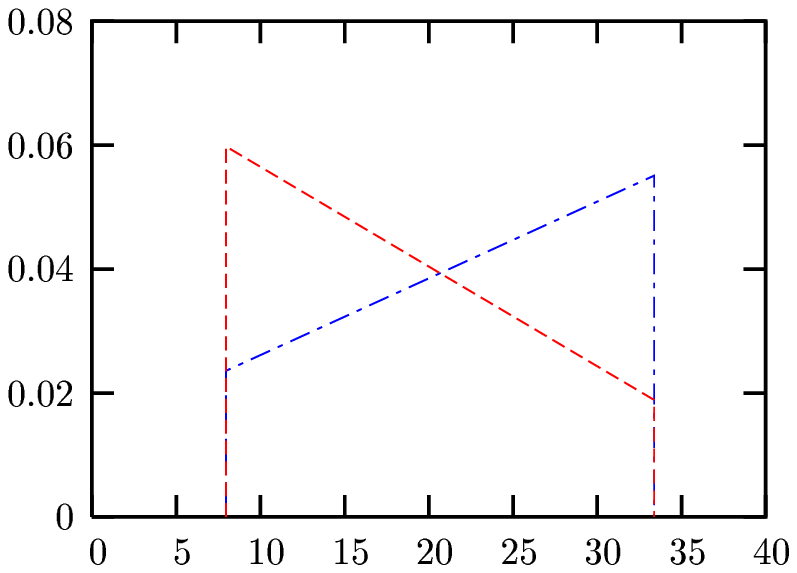}}
\put(1.4,3.85){$ \frac{1}{\sigma_\ell}\frac{d\sigma_\ell}{d E_\ell}
	{\scriptstyle[ \rm{GeV}^{-1}]}$}
\put(8.68,-.15){$ \scriptstyle E_\ell[\rm{GeV}]$}
\put(5.65,3.4){$  \textcolor{red}{\ell^-} $}
\put(8.3,3.2){$ \textcolor{blue}{\ell^+} $}
\end{picture}
\caption{\small 
	Normalized energy distributions of the lepton for the process
	$\mu^+\mu^-\to\chp\chm$ and  decay 
	$\chp\to\ell^+ \tilde\nu_\ell$ (dot-dashed) or
	$\chm\to\ell^-\tilde\nu_\ell^\ast$ (dashed), 
	for $\ell=e,\mu$, 
	with $\sqrt{s}=500$~GeV 
	and longitudinal beam polarizations 
	$\mathcal{P}_{+}=\mathcal{P}_{-}=-0.3$.
	The MSSM parameters are given in Table~\ref{scenarios1}. 
	The shown distributions have asymmetries
	$\mathcal{A}_{\ell^+}=0.2$ and
	$\mathcal{A}_{\ell^-}=-0.26$, see~(\ref{apoltot}).
}
\label{fig:edist.mu-500gm200}
\end{figure}

The energy distribution of the decay particle $\lambda^\pm$
is now given by 
\begin{equation}
\frac{d\sigma_{\lambda^\pm}}{dE_\lambda} =
	\frac{\sigma_\lambda}
		{2\Delta_\lambda}\left[ 1 + \,
						\eta_{\lambda^\pm}\frac{\bar{\Sigma}_P^{3}}{\bar{P}} 
	\frac{(E_\lambda - \bar{E_\lambda})}{\Delta_\lambda} \right], 
\label{edist2}
\end{equation}
where  we have defined averages
over the chargino production angles in the CMS by
\begin{eqnarray}
	\bar{P} 		= \frac{1}{4\pi}\int 	P 		d\Omega_{\chi^{\pm}},
\qquad
	\bar{\Sigma}^{3}_P	= \frac{1}{4\pi}\int {\Sigma}^{3}_P
	d\Omega_{\chi^{\pm}}. 
\label{eq:sigmabar}
\end{eqnarray}
Two examples of energy distributions of the decay particles
$\ell^+$ and $\ell^-$, for $\ell=e,\mu$, 
are shown in Fig.~\ref{fig:edist.mu-500gm200}.
One can see the linear dependence of the distributions 
on the lepton energy. The slope of the curves is proportional 
to the longitudinal chargino polarization.
Note that the energy distribution might be difficult to
measure for a small chargino-sneutrino mass difference,
since the energy span of the observed lepton is proportional to
the difference of their squared masses, see (\ref{edif}).

\subsection{Asymmetries of the energy distribution}

For the cross section $\sigma_{\lambda^\pm}$ of chargino 
production~(\ref{production}) with subsequent two body decay of one 
chargino into a lepton and a sneutrino~(\ref{decay1}) or 
into a $W$ boson and a neutralino~(\ref{decay2}), 
we define the asymmetries $\mathcal{A}_{{\lambda^+}}$ 
and $\mathcal{A}_{{\lambda^-}}$ for the charge conjugated processes
\begin{eqnarray}
	\mathcal{A}_{{\lambda^\pm}} &=& 
	\frac{\sigma_{\lambda^\pm}(E_\lambda>\bar{E}_\lambda)-\sigma_{\lambda^\pm}(E_\lambda<\bar{E}_\lambda)}{\sigma_{\lambda^\pm}(E_\lambda>\bar{E}_\lambda)+\sigma_{\lambda^\pm}(E_\lambda<\bar{E}_\lambda)}, \qquad \lambda=e,\mu,\tau,W.
\label{apoltot}
\end{eqnarray}
Using the formula for the energy distribution of the decay particle
$\lambda^\pm$~(\ref{edist2}), we find that the asymmetries are
proportional to the averaged longitudinal chargino polarization 
\begin{eqnarray}
\mathcal{A}_{{\lambda^\pm}} &=& \frac{1}{2}
		\eta_{\lambda^\pm}
	\frac{\bar{\Sigma}_P^{3}}{\bar{P}}.
	\label{apoltot2}
\end{eqnarray}
In order to separate the resonant contributions of the 
Higgs exchange channels to  $\bar{\Sigma}_P^3$ from those 
of the continuum contributions, see~(\ref{contributions})
and~(\ref{eq:sigmabar}), we use their different dependence 
on the chargino charge and on the beam polarizations. 
Therefore it is useful to discuss the production of equal and 
unequal charginos separately.

\subsubsection{Production of equal charginos}

If equal charginos are produced,
$\mu^+  \mu^-\to \tilde\chi_j^{+} \tilde\chi_j^{-}$,
the resonant contributions $\Sigma_{r}^{3\, (HA)}$
are independent of the chargino charge. 
The continuum contributions $\Sigma_{cont}^3$, however,
differ by a sign for charginos with 
positive or negative charge, see~(\ref{Ccontdependence}),
and are thus eliminated in the numerator of the charge asymmetries 
{\it }
\begin{eqnarray}
	\mathcal{A}_{\lambda}^{C}&=&
\frac{1}{2}[\mathcal{A}_{\lambda^+}-\mathcal{A}_{\lambda^-}]
\label{achargeA}\\
	&=&
\frac{1}{2}
		\eta_{\lambda^+}
	\frac{\Sigma_{r}^{3\, (HA)}}{\bar{P}},
	\qquad \lambda=e,\mu,\tau,W,
\label{acharge}
\end{eqnarray}
with $\bar\Sigma_r^{3\, (HA)}=\Sigma_r^{3\, (HA)}$,
see~(\ref{eq:sigmabar}).
The resonant contributions can also be isolated from the continuum 
contributions by taking into account their different 
dependence on the beam polarizations 
for $\mathcal{P}\equiv\mathcal{P}_+ =\mathcal{P}_-$, given 
in~(\ref{PPrdependence}), (\ref{PSigmaHArdependence}). 
Then the invariant continuum contributions 
are eliminated in the polarization asymmetries
\begin{eqnarray}
	\mathcal{A}_{{\lambda^\pm}}^{pol}&=&
\frac{1}{2}[\mathcal{A}_{{\lambda^\pm}}(\mathcal{P})
	-\mathcal{A}_{{\lambda^\pm}}(-\mathcal{P})]
\label{apolA}\\
	&=&
\frac{1}{2}
		\eta_{\lambda^\pm}
	\frac{\Sigma_r^{3\, (HA)}(\mathcal{P})}
	{\bar{P}}.
\label{apol}
\end{eqnarray}
Since $\Sigma_r^{3\, (HA)}$~(\ref{sigmacpconsHA}) 
describes 
the interference of the $H$ and $A$ exchange amplitudes, 
nonvanishing asymmetries $\mathcal{A}_{\lambda}^{C}$ and
$\mathcal{A}_{{\lambda^\pm}}^{pol}$ are a clear indication 
of nearly degenerate scalar resonances with opposite $CP$ 
quantum numbers in the production of equal charginos.

\subsubsection{Production of $\chpm_1\chmp_2$}

The asymmetries 
$\mathcal{A}_{\lambda}^{C}$ (\ref{acharge}) 
and $\mathcal{A}_{{\lambda^\pm}}^{pol}$  (\ref{apol})
have to be generalized for the production of unequal charginos,
$\chmp_1\chpm_2$,
since the coefficient $P_r^{(HA)}$~(\ref{pcpconsHA}) does not vanish.
For either the decay of $\chpm_1$ or the decay of $\chpm_2$
we define the generalized charge asymmetry 
\begin{eqnarray}
	\tilde{\mathcal{A}}_{{\lambda}}^C &=& 
	\frac{
	\sigma^{>}_{\lambda^+}-\sigma^{<}_{\lambda^+}
	-
	\sigma^{>}_{\lambda^-}+\sigma^{<}_{\lambda^-}
	}{
	\sigma^{>}_{\lambda^+}+\sigma^{<}_{\lambda^+}
	+
	\sigma^{>}_{\lambda^-}+\sigma^{<}_{\lambda^-}
	}, 
	\qquad \lambda=e,\mu,\tau,W,
\label{aCtot}
\end{eqnarray}
with the short hand notation
$\sigma^{>}_{\lambda^\pm} = \sigma_{\lambda^\pm}(E_\lambda>\bar{E}_\lambda)$
and 
$\sigma^{<}_{\lambda^\pm} = \sigma_{\lambda^\pm}(E_\lambda<\bar{E}_\lambda)$.
Using the definition of the energy distribution~(\ref{edist2}) and the
chargino charge transformation properties of the
coefficients $P$ and $\Sigma_P^3$, 
(\ref{Ccontdependence})-(\ref{Cresdependence}),
the resonant contributions can be separated, 
in analogy to~(\ref{acharge}),
\begin{eqnarray}
\tilde{\mathcal{A}}_{{\lambda}}^C &=& 
	\frac{1}{2}
		\eta_{\lambda^+}
		\frac{\Sigma_r^{3\, (HA)}}
	{\bar{P}_{cont}+{P}_r^{(HH)}+{P}_r^{(AA)}},
		\label{aCtot2}
\end{eqnarray}
with $\bar{P}_r^{(\phi\phi)}={P}_r^{(\phi\phi)}$.
Analogously we define the generalized polarization asymmetry 
\begin{eqnarray}
	\tilde{\mathcal{A}}_{{\lambda^\pm}}^{pol}&=&
	\frac{
		\sigma^{>}_{\lambda^\pm}(\mathcal{P})-
		\sigma^{<}_{\lambda^\pm}(\mathcal{P})
		-\sigma^{>}_{\lambda^\pm}(\mathcal{-P})+
		\sigma^{<}_{\lambda^\pm}(\mathcal{-P})
	}{
	\sigma^{>}_{\lambda^\pm}(\mathcal{P})+
	\sigma^{<}_{\lambda^\pm}(\mathcal{P})
	+
	\sigma^{>}_{\lambda^\pm}(\mathcal{-P})+
	\sigma^{<}_{\lambda^\pm}(\mathcal{-P})
	}
\label{apolijA}\\
	&=&
\frac{1}{2}
		\eta_{\lambda^\pm}
	\frac{\Sigma_r^{3\, (HA)}(\mathcal{P})}
{\bar{P}_{cont}+{P}_r^{(HH)}+{P}_r^{(AA)}},
	\qquad \lambda=e,\mu,\tau,W,
\label{apolijB}
\end{eqnarray}
for equal beam polarizations $\mathcal{P}$.
For the production of  equal charginos 
these asymmetries reduce to their equivalents 
$\mathcal{A}_{\lambda}^{C}$ and ${\mathcal{A}}_{{\lambda^\pm}}^{pol}$,
defined in~(\ref{acharge}) and (\ref{apol}), respectively.

Moreover we define the production asymmetry of the chargino 
cross sections  
\begin{eqnarray}
\mathcal{A}^C_{pr\!od} = \frac{\sigma(\charginoplus_1\charginominus_2)-
									\sigma(\charginoplus_2\charginominus_1)}
									{\sigma(\charginoplus_1\charginominus_2)+
									\sigma(\charginoplus_2\charginominus_1)}
						= \frac{P_r^{(HA)}}
								{\bar{P}_{cont}+P_r^{(HH)}+P_r^{(AA)}},
\label{aprodch1ch2}
\end{eqnarray}
which is sensitive to the interference of the $H$ and $A$ channels
due to the parity violating Higgs-chargino couplings.

\subsubsection{Statistical significances}

We define the statistical significance of the asymmetries
$\mathcal{A}_{{\lambda^\pm}}$ by
\begin{eqnarray}
	\mathcal{S}_{\lambda^\pm}
	&=& |\mathcal{A}_{\lambda^\pm}|
	\sqrt{\sigma(\mu^+\mu^-\to\chmp_i\chpm_j)
		{\rm{BR}}(\chpm_j\to\lambda^\pm\tilde N_\lambda)	
{	\mathcal{L}_{ef\!f}	}},\label{significance}
\end{eqnarray}
with $\lambda=\ell$ or $W$ and $\tilde N_\lambda$ the associated 
sneutrino or neutralino, respectively. 
Further the effective integrated luminosity
$\mathcal{L}_{ef\!f} =\eps_\lambda \mathcal{L}$
depends on the detection efficiency $\eps_\lambda$ of leptons or 
$W$ bosons in the processes
$\chpm_j\to\ell^\pm~\sneutrino_\ell^{(\ast)}$ or 
$\chpm_j\to W^\pm~\neutralino_k$, respectively.
The statistical significance for the charge asymmetry 
$\mathcal{A}_{\lambda}^{C}$ is given by
\begin{eqnarray}
        \mathcal{S}_{\lambda}^{C}
        &=& |\mathcal{A}_{\lambda}^{C}| \sqrt{2\, 
\sigma(\mu^+\mu^-\to\chmi\chpj)
{\rm{BR}}(\chpj\to\lambda^+\tilde N_\lambda) 
					 {       \mathcal{L}_{ef\!f}     }},
	\label{significancechar}
\end{eqnarray}
which follows from~(\ref{achargeA}).
Assuming that  
$\mathcal{A}_{{\lambda^\pm}}(\mathcal{P})$
and
$\mathcal{A}_{{\lambda^\pm}}(-\mathcal{P})$
are both obtained with the same integrated luminosity 
$\mathcal{L}$, we define the statistical significance 
for the polarization asymmetry 
$\mathcal{A}_{{\lambda^\pm}}^{pol}$
by
\begin{eqnarray}
        \mathcal{S}_{\lambda^\pm}^{pol}
        &=& |\mathcal{A}_{\lambda^\pm}^{pol}| \sqrt{2\, 
\sigma(\mu^+\mu^-\to\chmp_i\chpm_j)
{\rm{BR}}(\chpm_j\to\lambda^\pm\tilde N_\lambda) 
					 {       \mathcal{L}_{ef\!f}     }},
	\label{significancepol}
\end{eqnarray}
which follows from~(\ref{apolA}).
For the production asymmetry $\mathcal{A}^C_{pr\!od}$~(\ref{aprodch1ch2})
we define the significance
\begin{eqnarray}
        \mathcal{S}^C_{pr\!od}
        &=& |\mathcal{A}^C_{pr\!od}| \sqrt{ 
			  [\sigma(\charginoplus_1\charginominus_2) + 
			   \sigma(\charginoplus_2\charginominus_1)]
        {    \mathcal{L}_{ef\!f}^{pr\!od}     }},
	\label{significanceprod}
\end{eqnarray}
with ${\mathcal{L}_{ef\!f}^{pr\!od}}$ the effective integrated 
luminosity for chargino production.

\section{Determination of the Higgs-chargino couplings
  \label{Determination of the Higgs-chargino couplings}}
In the previous section we have shown that the 
coefficient $\Sigma_{r}^3 $~(\ref{sigmacpconsHA})
of the longitudinal chargino polarization 
is sensitive to the interference of the $H$ and $A$ Higgs bosons.
Their interference determines the 
sign $\gamma$ of the product of couplings
\begin{equation}
	\kappa=
	{\rm Im}\{c^{(H\mu)}c^{(A\mu)\ast}\}
	{\rm Im}\{c_{R}^{(H)}c_{R}^{(A)\ast}\}
	=\gamma\,|c^{(H\mu)}c^{(A\mu)}c_{R}^{(H)}c_{R}^{(A)}|, 
\label{kappa}
\end{equation}
which appears in 
\begin{equation}
	\Sigma_{r}^{3(HA)}  =  
	 2 g^4 \mathcal{P} \mbox{Re}\{\Delta(H)\Delta(A)^\ast\} 
	 {\rm Im}\{c^{(H\mu)}c^{(A\mu)\ast}\}
	 {\rm Im}\{c_{R}^{(H)}c_{R}^{(A)\ast}\}
	 s\sqrt{\lambda_{11}},
	\label{lalelu1}
\end{equation}
where we focus on the production of the lightest pair of charginos 
$\mu^+ \mu^-\to \tilde\chi_1^+\tilde\chi_1^-$
with equal muon beam polarizations 
$\mathcal{P}_+=\mathcal{P}_-\equiv \mathcal{P}$.
Since we assume $CP$ conservation, $\gamma$ can take the 
value $\pm1$ for interfering amplitudes of opposite $CP$ 
eigenvalues, and vanishes for interfering amplitudes 
with same $CP$ eigenvalues.
A measurement of $\gamma$ would thus be
a unique test of the $CP$ properties of the Higgs 
sector in the underlying supersymmetric model.

The coefficient $\Sigma_{r}^{3(HA)}$ can be obtained
from the chargino production cross section
\begin{eqnarray}
{\sigma}(\mu^+\mu^-\to\chp\chm) 
&=& \frac{\sqrt{\lambda_{11}}}{8\pi s^2} \bar P, 
\end{eqnarray}
and the charge asymmetry
$\mathcal{A}_{{\lambda}}^{C}$~(\ref{acharge})
\begin{equation}
\Sigma_{r}^{3(HA)}=
\frac{16\pi s^2}{\eta_{\lambda^+}\sqrt{\lambda_{11}}}\,
	\sigma(\mu^+\mu^-\to\chp\chm)\,
	\mathcal{A}_{{\lambda}}^{C}.
\label{lalelu2}
\end{equation}
Now the product of couplings $\kappa$ can be determined
by a comparison of~(\ref{lalelu2}) with~(\ref{lalelu1}). 
Alternatively, using the polarization asymmetry
$\mathcal{A}_{{\lambda^\pm}}^{pol}$~(\ref{apol}),
we find
\begin{equation}
\Sigma_{r}^{3(HA)}=
\frac{16\pi s^2}{\eta_{\lambda^\pm}\sqrt{\lambda_{11}}}\,
	\sigma(\mu^+\mu^-\to\chp\chm)\,
	\mathcal{A}_{{\lambda^\pm}}^{pol}.
\label{lalelu3}
\end{equation}

In addition, a measurement of the asymmetries 
$\mathcal{A}_{{\lambda}}^{C}$~(\ref{acharge})
or
$\mathcal{A}_{{\lambda^\pm}}^{pol}$~(\ref{apol})
allows the determination of the ratio
\begin{eqnarray}
	\frac{{\Sigma}_{r}^3}{{P_{r}}}
&=& 	\frac{\sigma(\mu^+\mu^-\to\tilde\chi_1^+\chi_1^-)}
	     {\sigma_{r}(\mu^+\mu^-\to\tilde\chi_1^+\chi_1^-)}
		\frac{2}{\eta_{\lambda^+}}
		  \mathcal{A}_{{\lambda}}^{C}
\label{lambdahaC}\\
&=& 	\frac{\sigma(\mu^+\mu^-\to\tilde\chi_1^+\chi_1^-)}
	     {\sigma_{r}(\mu^+\mu^-\to\tilde\chi_1^+\chi_1^-)}
		\frac{2}{\eta_{\lambda^\pm}}
		\mathcal{A}_{{\lambda^\pm}}^{pol},
\label{lambdaha}
\end{eqnarray}
using the charge or polarization asymmetry, respectively.
The resonant contributions 
\begin{eqnarray}
{\sigma}_{{r}}(\mu^+\mu^-\to\chp\chm) 
&=& \frac{\sqrt{\lambda_{11}}}{8\pi s^2} P_r, 
\quad {\rm with} \quad P_{r} =\bar P_{r}, 
\end{eqnarray}
to the cross section  
can be obtained by subtracting the continuum contributions. 
The latter can be estimated by extrapolating 
the production line shape below and 
above the resonance 
region~\cite{Fraas:2003cx}. Uncertainties due to detection efficiencies
of the chargino decay products cancel out in the ratio  
\begin{eqnarray}
	\frac{\sigma_{r}(\mu^+\mu^-\to\tilde\chi_1^+\chi_1^-)}
	     {\sigma(\mu^+\mu^-\to\tilde\chi_1^+\chi_1^-)} &=& 		
				\frac{P_{r}}{\bar{P}}.
\label{Pint}
\end{eqnarray}
After inserting the expressions of 
$\Sigma_{r}^{3(HA)}$~(\ref{lalelu1}) and
$P_r$~(\ref{pcpconsphiphi}) 
we obtain 
\begin{eqnarray}
\frac{{\Sigma}_{r}^3}{{P_{r}}}
&=&
 \frac{2\mathcal{P}}{1+\mathcal{P}^2}
 \,\frac{
	 2\gamma\, \mbox{Re}\{\Delta(H)\Delta^*(A)\} \sqrt{s^+s}
	}
	{ r \,|\Delta{(H)}|^2 \,s^+ + r^{-1} \, |\Delta{(A)}|^2\, s },
\label{apol1}
\label{apol2}
\end{eqnarray}
with
\begin{eqnarray}
s^+&=& s -4{m}_{\chi_1^\pm}^2=\frac{\lambda_{11}}{s},
\label{s_pm}\\
r&=&\frac{|c^{(H\mu)}c_R^{(H)}|}{| c^{(A\mu)} c_R^{(A)}|}.
\end{eqnarray}
It is now possible to solve~(\ref{apol2}) for $r$ as well as for $\gamma$. 

For our analysis we have assumed that the masses and widths of the 
Higgs resonances $H$ and $A$ can be measured. The resonance parameters 
of nearly degenerate Higgs bosons with different $CP$ quantum numbers 
may e.g. be determined by using transverse beam polarizations, 
which  enhances or suppresses the Higgs exchange channels 
depending on their $CP$ quantum numbers~\cite{poltransv}.

Note that $\gamma$~(\ref{kappa}) 
can only be determined by measuring the 
charge or polarization asymmetries
$\mathcal{A}_{\lambda}^{C}$ and $\mathcal{A}_{{\lambda^\pm}}^{pol}$,
which are sensitive to the $H$--$A$ interference channels. 
A determination of $\gamma$ from a measurement of the cross section 
$\sigma(\mu^+\mu^-\to\chp\chm)$ is not possible,
since it contains contributions from pure $H$ or $A$ exchange only.

\section{Numerical results
  \label{Numerical results}}

We analyze numerically the 
charge asymmetry $\mathcal{A}_{\ell}^C$~(\ref{achargeA}) 
of the lepton energy distribution 
for the production of equal charginos
$\mu^+\mu^-\to\tilde\chi^+_1\tilde\chi^-_1$
in Section \ref{num:Production.of.chi1chi1},
and the cross section asymmetry 
$\mathcal{A}^C_{pr\!od} $~(\ref{aprodch1ch2})
for the production of different charginos
in Section \ref{num:Production.of.chi1chi2}.
The feasibility of measuring the asymmetries depends also
on the corresponding production cross sections 
which we discuss in our scenarios.
For the calculation of the Higgs masses and widths we use
the program HDECAY~\cite{HDECAY}.
For the calculation of the branching ratios 
and widths 
of the decaying charginos we include the two-body decays
\begin{eqnarray}\label{chardecaymodes}
	\tilde\chi^\pm_1 &\to& 
	e^\pm\tilde\nu_{e},~
	\mu^\pm\tilde\nu_{\mu},~
	\tau^\pm\tilde\nu_{\tau},~
	\tilde e_{L}^\pm\nu_{e},~
	\tilde\mu_{L}^\pm\nu_{\mu},~
	\tilde\tau_{1,2}^\pm\nu_{\tau},~
	W^\pm\tilde\chi^0_n,
\end{eqnarray}
and neglect three-body decays. 
In order to reduce the number of parameters, we assume  
GUT relations for the gaugino mass parameters, related by  
$M_1=5/3 \, M_2\tan^2\theta_W $, and for the slepton masses, related to 
the scalar mass parameter $m_0$ at the GUT scale by the
approximate renormalization group equations \cite{Hall:zn}
\begin{eqnarray}
	m_{\tilde\ell_R}^2 &=& m_0^2 +0.23 M_2^2
	-m_Z^2\cos 2 \beta \sin^2 \theta_W,\\ 
	m_{\tilde\ell_L  }^2 &=& m_0^2 +0.79 M_2^2
	+m_Z^2\cos 2 \beta(-\frac{1}{2}+ \sin^2 \theta_W),\\ 
	m_{\tilde\nu_{\ell}  }^2 &=& m_0^2 +0.79 M_2^2 +
	\frac{1}{2}m_Z^2\cos 2 \beta.
\label{msll}
\end{eqnarray}
In the stau sector we fix the trilinear scalar coupling
parameter $A_{\tau}=250$~GeV.

\subsection{Production of $\charginoplus_1\charginominus_1$
	\label{num:Production.of.chi1chi1}}

In the following subsections we study the dependence of the
asymmetries and cross sections on the MSSM parameters 
$\mu$, $M_2$, $\tan\beta$ and $m_A$, 
as well as on the center of mass energy  $\sqrt{s}$. 

\subsubsection{  $\mu$ and $M_2$ dependence}

\begin{figure}[t]
\centering
\begin{picture}(16,8.)
\put(-2.2,-14.6){\includegraphics{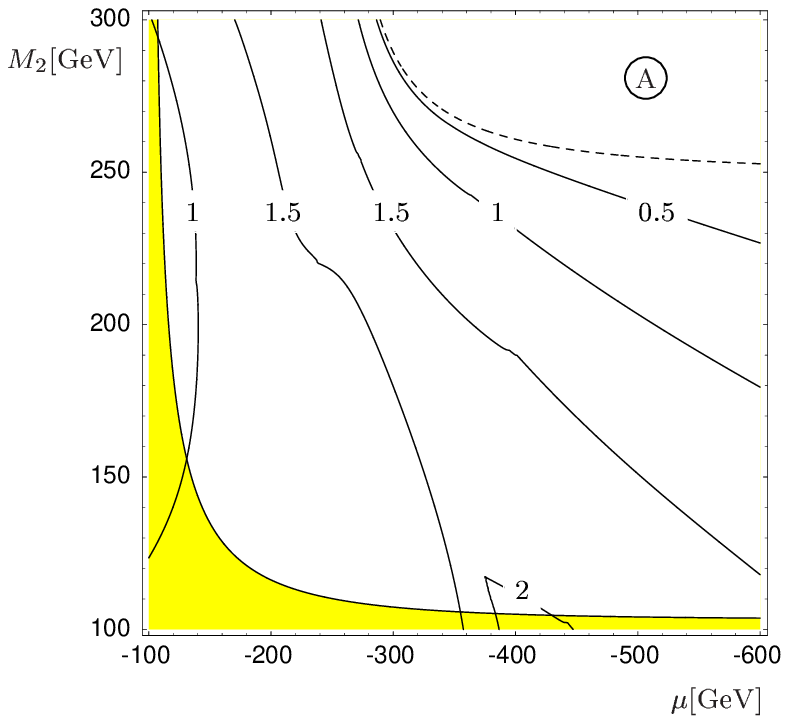}}
\put(2.15,7.6){$\sigma(\mu^+\mu^-\to\charginoplus_1\charginominus_1)$ in $\rm{pb}$}
\put(1,0.3){ (a)}
\put(5.65,-14.6){\includegraphics{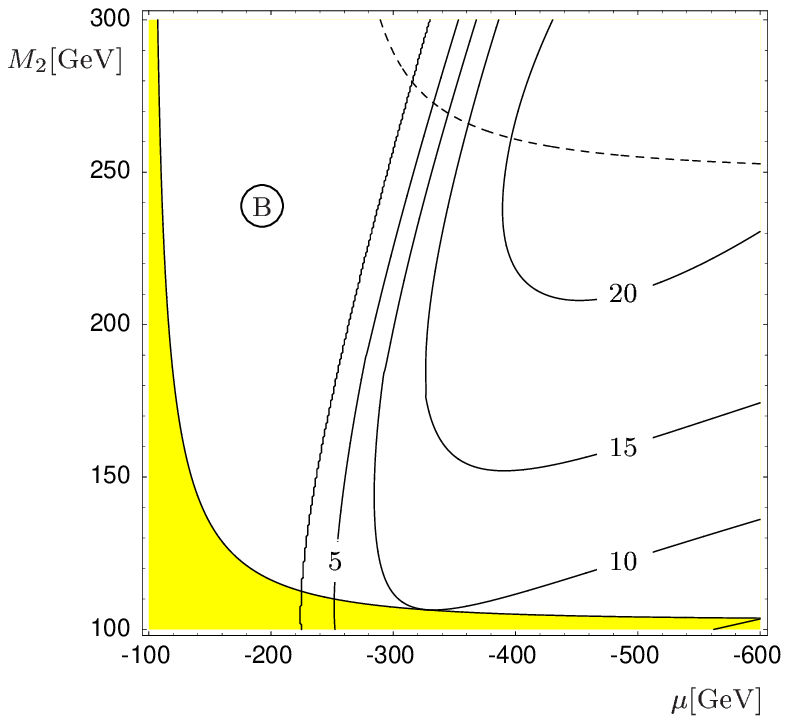}}
\put(10.35,7.6){${\rm{BR}}(\charginoplus_1\to e^+\sneutrino_e)$ in \%}
\put(8.9,0.3){ (b)}
\end{picture}
\begin{picture}(16,8.)
\put(-3.05,-14.8){\includegraphics{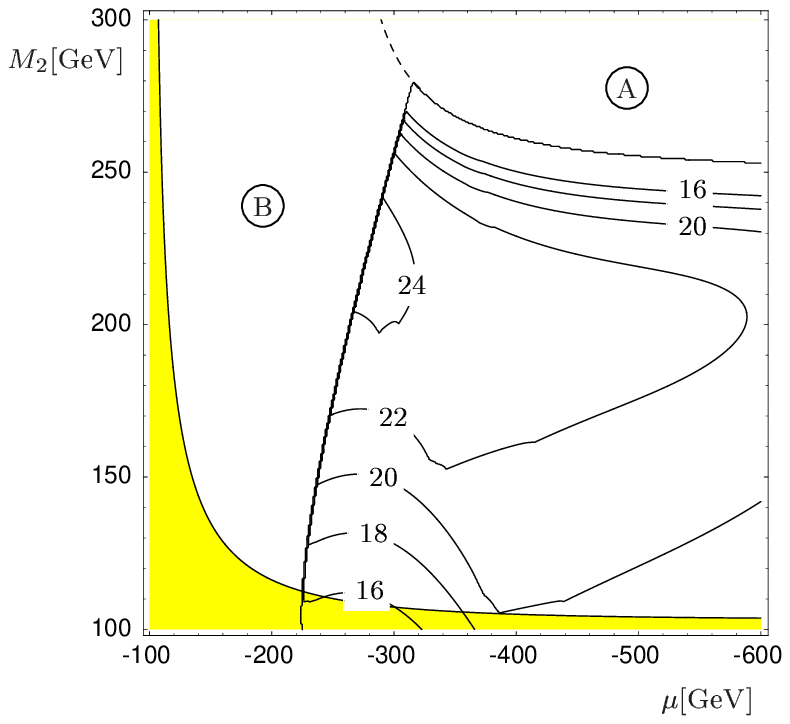}}
\put(2.45,7.4){Asymmetry $\mathcal{A}^C_e$ in \%}
\put(1,0.1){ (c)}
\put(4.7,-14.8){\includegraphics{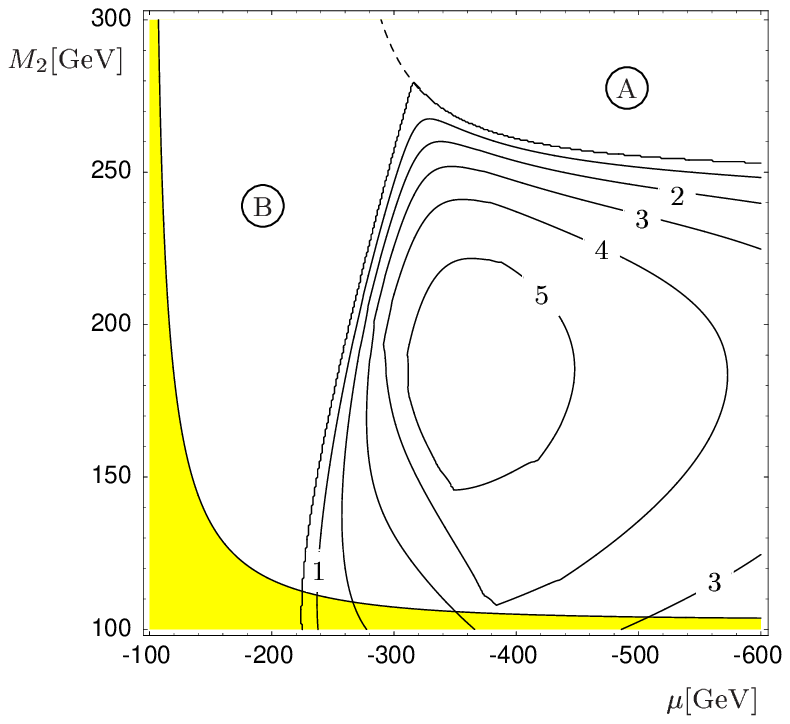}}
\put(10.65,7.4){Significance $\mathcal{S}^C_e$}
\put(8.9,0.1){ (d)}
\end{picture}
\caption{\small 
	$ \mu^+\mu^-\to\chargino_1\charginominus_1$,
	$\charginoplus_1\to e^+\sneutrino_e$. 
	Contour lines of the cross section 
	$\sigma(\mu^+\mu^-\to\chargino_1\charginominus_1)$ {\bf{(a)}}, 
	the branching ratio ${\rm{BR}}(\charginoplus_1\to e^+\sneutrino_e)$ {\bf{(b)}},
	the charge asymmetry $\mathcal{A}^C_e$ {\bf{(c)}}
	and the significance $\mathcal{S}^C_e$  for $\mathcal{L}_{ef\!f}=1~\rm{fb}^{-1}$ {\bf{(d)}}
	in the $\mu$--$M_2$ plane for $m_A=500$~GeV, $\tan \beta=10$, 
	$m_0=70$~GeV, $\sqrt{s}=500$~GeV and 
	longitudinal beam polarizations 
	$\mathcal{P}_{+}=\mathcal{P}_{-}=-0.3$.
	The dashed line indicates the kinematical limit
	$2m_{\chi_1^\pm}=\sqrt{s}$.
	The area A (B) is kinematically forbidden by
	$2m_{\chi_1^\pm}>\sqrt{s}$
	$(m_{\tilde\nu_e}> m_{\chi^\pm_1})$.
	The shaded  area is excluded by $m_{\chi_1^\pm}<103$~GeV.
}
\label{fig:contours}
\end{figure}
%
In Fig.~\ref{fig:contours}a 
we show the contour lines of the
chargino production cross section 
$\sigma(\mu^+\mu^-\to\tilde\chi^+_1\tilde\chi^-_1 )$
in the $\mu$--$M_2$ plane for $\sqrt{s}=m_A$
and beam polarizations
$\mathcal{P}_{+}=\mathcal{P}_{-}=-0.3$,
with $m_A=500$~GeV, $\tan \beta=10$ and $m_0=70$~GeV.
At $\sqrt{s}=m_A\approx m_H$ the production cross section is close to 
its peak value, since the two Higgs resonances 
are nearly degenerate. The main contributions to the cross section, 
which reaches up to $2$~pb, are from the resonant ones. 
For increasing values of $|\mu|$ the couplings of both $H$ and $A$ to 
the charginos decrease, leading to smaller resonant contributions. 
The continuum contributions from $\gamma$, $Z$ and $\tilde\nu_\mu$ exchange
reach $0.5$~pb at most.

We show contour lines of the chargino branching ratio 
${\rm{BR}}(\tilde\chi^+_1\to e^+\tilde\nu_e)$
in the $\mu$--$M_2$ plane in  Fig.~\ref{fig:contours}b, where also 
the allowed region for the chargino two-body decay 
$\tilde\chi^+_1\to e^+\tilde\nu_e$ is indicated. 
The sneutrinos are rather light for $m_0=70$~GeV, 
such that this chargino decay mode is open for  
$|\mu| \gsim 200$~GeV and reaches values of up to 20\%.

For the chargino decay into an electron
$\tilde\chi^\pm_1\to e^\pm\tilde\nu_e^{(\ast)}$, 
we show in Fig.~\ref{fig:contours}c 
contour lines of the charge asymmetry  
$\mathcal{A}_{e}^C$~(\ref{achargeA})
which reaches values of up to 24\%.
The asymmetry 
depends only weakly on 
the character of chargino mixing, 
since $\mathcal{A}_{e}^C$ is proportional to a ratio of the couplings,
see (\ref{lambdahaC}) and (\ref{apol2}). In the ideal case of
maximal $H$-$A$ interference and vanishing continuum contributions,
the asymmetry could reach its maximum absolute value of 
$|\mathcal{P}_{+} +\mathcal{P}_{-}|/(1+\mathcal{P}_{+} \mathcal{P}_{-})/2   \approx 28\%$, 
as follows from (\ref{acharge})
for 
$\mathcal{P}_{+}=\mathcal{P}_{-} =-0.3$. 
Thus the shown values of $\mathcal{A}_{e}^C$ in Fig.~\ref{fig:contours}c
are large, since the amplitudes of the interfering 
$H$ and $A$ Higgs bosons 
are roughly of the same 
magnitude in the resonance region $\sqrt{s}=m_A$. 
Near the production threshold $\sqrt{s}=2m_{\chi_1^\pm}$ 
the asymmetry decreases due to the p-wave 
suppression of the $CP$ even scalar exchange amplitude.



In Fig.~\ref{fig:contours}d we show the contour lines of
the significance $\mathcal{S}_{e}^{C}$~(\ref{significancechar})
for an integrated effective luminosity 
$\mathcal{L}_{ef\!f}=1~\rm{fb}^{-1}$.
Due to the large asymmetry $\mathcal{A}_{e}^C$ and cross section 
$\sigma(\mu^+\mu^-\to\tilde\chi^+_1\tilde\chi^-_1) \times
{\rm BR}(\tilde\chi^+_1\to e^+\tilde\nu_e)$
for chargino production and subsequent decay, 
$\mathcal{A}_{e}^C$ can be measured with a significance
$\mathcal{S}_{e}^{C}>1$ for a luminosity 
	$\mathcal{L}_{ef\!f}=\mathcal{O}(\rm{fb}^{-1})$.
The same values of the significance are obtained for 
the muonic chargino decay mode $\tilde\chi^+_1\to \mu^+\tilde\nu_\mu$.

\subsubsection{ $\sqrt{s}$ dependence}

In order to study the dependence of
the asymmetries and the chargino
production cross sections on the center of mass energy,
we choose a representative point in the $\mu$--$M_2$ plane 
with $\mu=-500$~GeV and $M_2=200$~GeV.
The parameters and resulting Higgs masses and widths
for this point, called scenario {\bf A}, are given in  Table~\ref{scenarios1}.  
  For the calculation of the branching ratios
  we include mixing in the stau sector,
  see e.g.~\cite{HK,kernreiter:2002}.
  Note that
  BR$(\chp\to W^+\tilde\chi^0_1)<0.3\%$
  due to the small $\chp$-$W^+$-$\neutralino_1$ coupling in the gaugino 
  scenario {\bf A}, 
  and  BR$(\chp\to \tilde e_L^+\nu_e) <0.01\%$
  due to kinematical reasons since
  $m_{\chi^\pm_1} \approx  m_{\tilde e_L}$.
\begin{table}
\renewcommand{\arraystretch}{1.2}
\caption{ Scenario {\bf A} for $\mu^+\mu^-\to\chp\chm$.}
\begin{center}
	\begin{tabular}{|l|c|c|c|}
\hline
$\tan\beta=10 $ & $m_A=500$ GeV	&	$m_{\chi^\pm_1}=197$ GeV 
&BR$(\chp\to e^+\sneutrino_e)=19\% $
\\
\hline
$\mu= -500$ {GeV} &  $\Gamma_A=1.41$ GeV	&	$m_{\chi^\pm_2}=514$ GeV
&BR$(\chp\to \mu^+\sneutrino_\mu)=19\% $
\\
\hline
$M_2= 200$ GeV	& $m_H=500.07$ GeV	&	$m_{\chi^0_1}=100$ GeV
&BR$(\chp\to \tau^+\tilde\nu_\tau)=19\% $
\\
\hline
$m_0=70$ GeV & $\Gamma_H=1.20 $ GeV & $m_{\tilde{\nu}_e}=180$ GeV
&BR$(\chp\to\tilde\tau_1^+\nu_\tau)=43\% $
\\
\hline
\end{tabular}
\end{center}
\renewcommand{\arraystretch}{1.0}
\label{scenarios1}
\end{table}

\begin{figure}[t]
\centering
\begin{picture}(16,6.7)
\put(-3.9,-10.6){\includegraphics{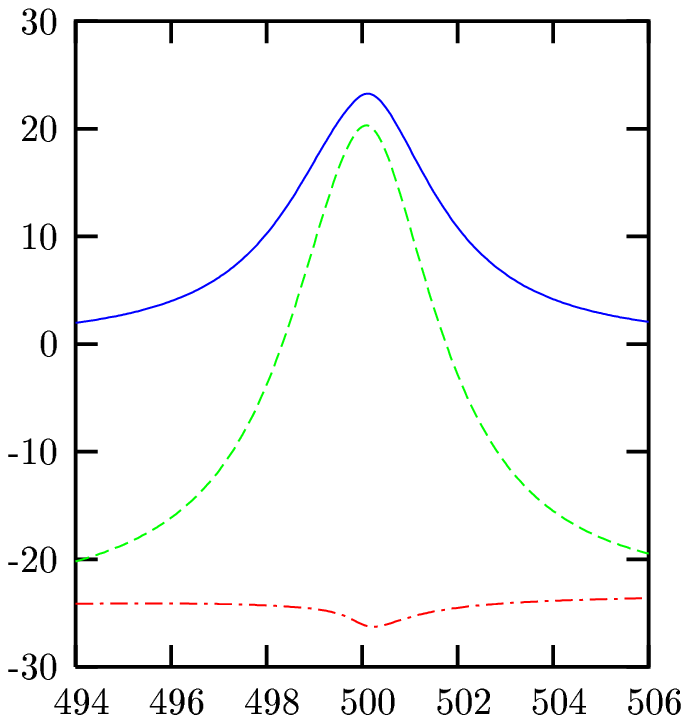}}
\put(0.6,-0.){ (a)}
\put(1.25,3.75){\small $ \mathcal{A}_e^C$}
\put(1.25,2.45){\small $ \mathcal{A}_{e^+}$}
\put(2.2,1.55){\small $ \mathcal{A}_{e^-}$}
\put(3.8,.15){$ \scriptstyle \sqrt{s}[\rm{GeV}]$}
\put(5.9,-10.6){\includegraphics{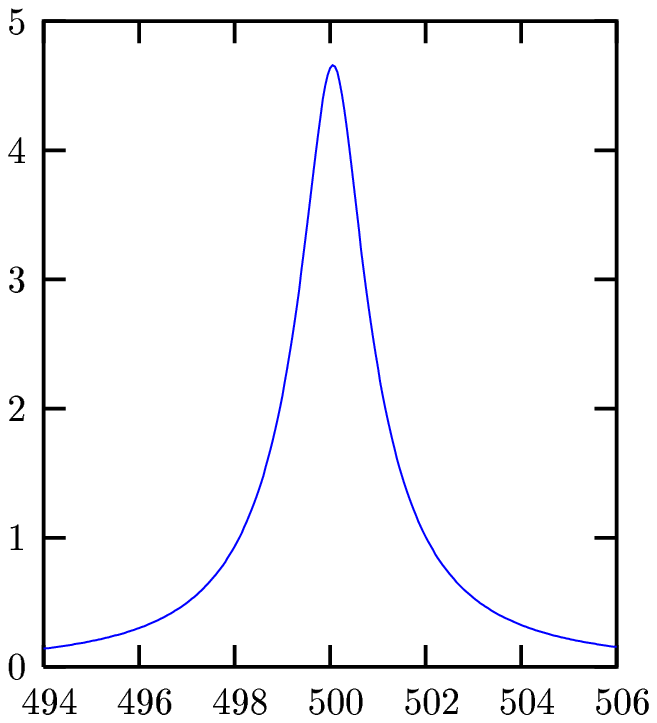}}
\put(1.0,-10.6){\includegraphics{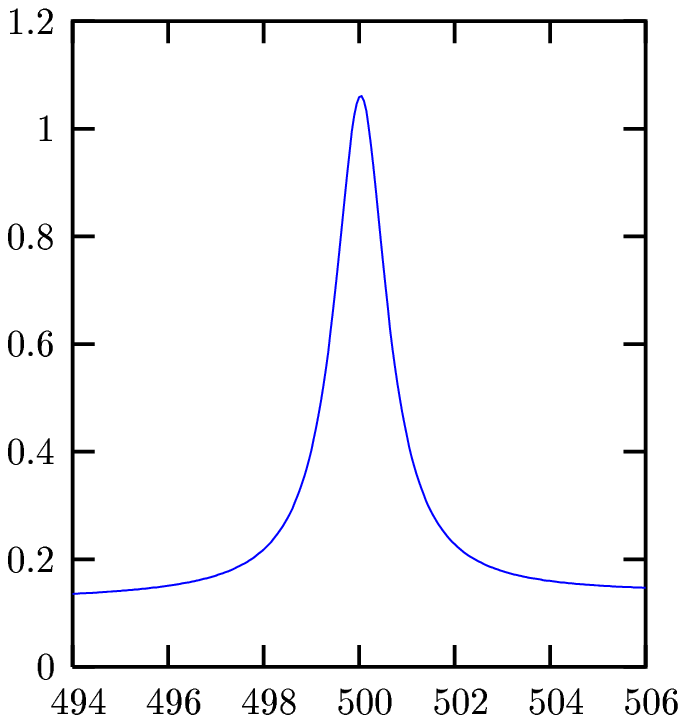}}
\put(5.5,0.){ (b)}
\put(10.4,0.){ (c)}
\put(8.7,.15){$ \scriptstyle \sqrt{s}[\rm{GeV}]$}
\put(13.6,.15){$ \scriptstyle \sqrt{s}[\rm{GeV}]$}
\put(1.,5.9){Asymmetries $\mathcal{A}$ in \%}
\put(5.6,5.9){$ \sigma(\mu^+\mu^-\to\chargino_1\charginominus_1)$ in $\rm{pb}$}
\put(11.3,5.9){Significance $\mathcal{S}_e^C$}
\end{picture}
\caption{\small 
	$ \mu^+\mu^-\to\chargino_1\charginominus_1$, 
	$\chpm_1\to e^\pm\sneutrino_e^{(\ast)}$. 
	Asymmetries {\bf{(a)}},
	chargino production cross section {\bf{(b)}} and
	significance for $\mathcal{L}_{ef\!f}=1~\rm{fb}^{-1}$ {\bf{(c)}},
	with longitudinal beam polarizations 
	$\mathcal{P}_{+}=\mathcal{P}_{-}=-0.3$
	for Scenario {\bf{A}}, given in Table~\ref{scenarios1}.
}
\label{fig:asymm.prod.B}
\end{figure}

In  Fig.~\ref{fig:asymm.prod.B}a we show 
the energy distribution asymmetry
$ \mathcal{A}_{e^+}$~(\ref{apoltot})
for the decay $\tilde\chi^+_1\to e^+\tilde\nu_e$,
and the asymmetry $\mathcal{A}_{e^-}$ for the charge conjugated process,
with longitudinal beam polarizations 
$\mathcal{P}_{+}=\mathcal{P}_{-}=-0.3$.
In addition we show the charge asymmetry
$\mathcal{A}_{e}^{C}=(\mathcal{A}_{e^+}-\mathcal{A}_{e^-})/2$, see~(\ref{achargeA}),
which reaches its maximal value of $23\%$ at $\sqrt{s}\approx m_A=500$~GeV.
Since the continuum contributions from
$\gamma$, $Z$ and $\tilde\nu_\mu$ exchange cancel out, 
$\mathcal{A}_{e}^{C}$ asymptotically vanishes far from the resonance region.
The $\sqrt{s}$ dependence of the chargino production 
cross section is shown in Fig.~\ref{fig:asymm.prod.B}b.
We show the corresponding statistical significance $\mathcal{S}_{e}^{C}$, 
defined in~(\ref{significancechar}), for an effective integrated 
luminosity $\mathcal{L}_{ef\!f}=1~\rm{fb}^{-1}$
in Fig.~\ref{fig:asymm.prod.B}c.

\subsubsection{ $m_A$ and $\tan\beta$ dependence}
\begin{table}
\renewcommand{\arraystretch}{1.2}
\caption{ Scenario {\bf B7} for $\mu^+\mu^-\to\chp\chm$,
chargino and slepton  parameters.}
\begin{center}
	\begin{tabular}{|l|c|c|}
\hline
$\tan\beta=7 $ 		&	$m_{\chi^\pm_1}=158$ GeV 
&BR$(\chp\to e^+\sneutrino_e)=22\% $
\\
\hline
$\mu= -400$~GeV		&	$m_{\chi^\pm_2}=417$ GeV
&BR$(\chp\to \mu^+\sneutrino_\mu)=22\% $
\\
\hline
$M_2= 160$~GeV		&	$m_{\chi^0_1}=81$ GeV\hspace{1mm}
&BR$(\chp\to \tau^+\tilde\nu_\tau)=22\% $
\\
\hline
$m_0=70$ GeV 		& $m_{\tilde{\nu}_e}=145$ GeV
&BR$(\chp\to\tilde\tau_1^+\nu_\tau)=33\% $
\\
\hline
\end{tabular}
\end{center}
\renewcommand{\arraystretch}{1.0}
\label{scenario.B7}
\end{table}
\begin{table}
\renewcommand{\arraystretch}{1.2}
\caption{Scenarios {\bf B7},  {\bf B7'} and {\bf B7''},
	Higgs sector parameters.
}
\begin{center}
	\begin{tabular}{|r|ccc|}
\hline
& {\bf B7} & {\bf B7'} &  {\bf B7''} 
\\
\hline
\hline
$m_A[\rm{GeV}]$ & $350$ 	& $400$ 	& $500$
\\
\hline
$m_H[\rm{GeV}]$ &  $350.7$	& $400.6$	&$500.4$
\\
\hline
 $\Gamma_A[\rm{GeV}] $		&  $0.56$	& $1.00$	& $1.4$
 \\
\hline
$\Gamma_H[\rm{GeV}] $		&  $0.43$	& $0.65$	& $1.1$
\\
\hline
\end{tabular}
\end{center}
\renewcommand{\arraystretch}{1.0}
\label{scenario.B7.Higgs}
\end{table}

\begin{figure}[t]
\centering
\begin{picture}(16,6.7)
\put(-3.9,-10.6){\includegraphics{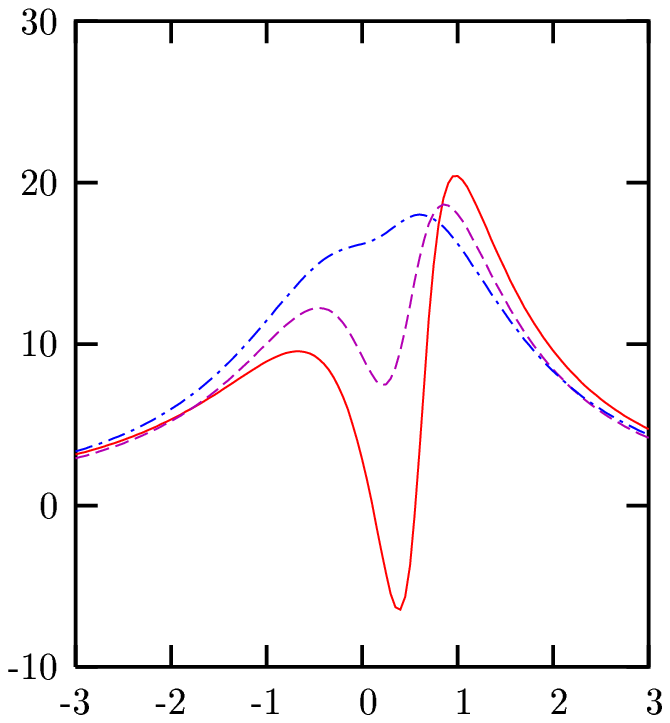}}
\put(5.9,-10.6){\includegraphics{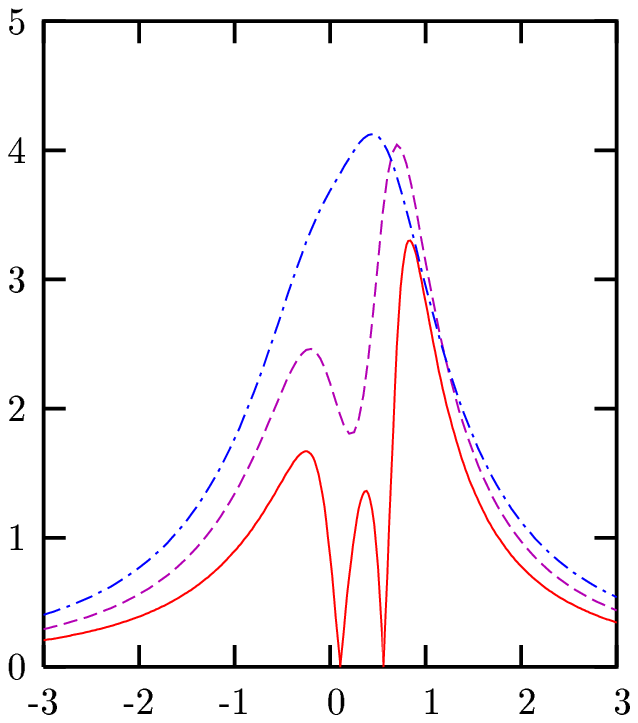}}
\put(1.0,-10.6){\includegraphics{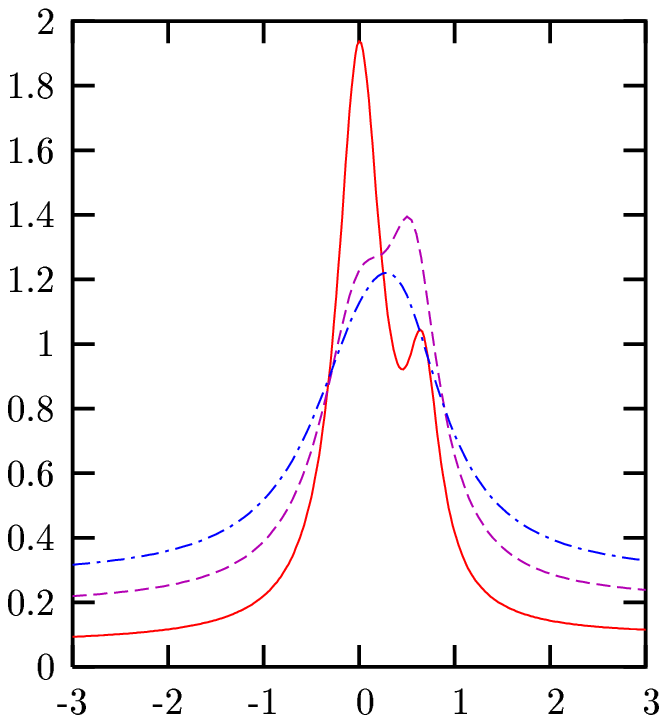}}
\put(0.6,-0.){ (a)}
\put(5.5,0.){ (b)}
\put(10.4,0.){ (c)}
\put(3.1,.15){$ \scriptstyle \sqrt{s}-m_A[\rm{GeV}]$}
\put(8,.15){$ \scriptstyle \sqrt{s}-m_A[\rm{GeV}]$}
\put(12.9,.15){$ \scriptstyle \sqrt{s}- m_A[\rm{GeV}]$}
\put(1.,5.9){Asymmetry $\mathcal{A}_{e}^{C}$ in \%}
\put(5.6,5.9){$ \sigma(\mu^+\mu^-\to\chargino_1\charginominus_1)$ in $\rm{pb}$}
\put(11.3,5.9){Significance $\mathcal{S}_e^C$}
\end{picture}
\caption{\small 
	$ \mu^+\mu^-\to\chargino_1\charginominus_1$, 
	$\chpm_1\to e^\pm\sneutrino_e^{(\ast)} $.
	Asymmetry $\mathcal{A}_{e}^{C}$ {\bf (a)},
	cross section $\sigma(\mu^+\mu^-\to\chargino_1\charginominus_1)$ {\bf (b)}
	and significance $\mathcal{S}_e^C$ for
	$\mathcal{L}_{ef\!f}=1~\rm{fb}^{-1}$ {\bf (c)} 
	for scenarios {\bf B7} (solid), {\bf B7'} (dashed) and 
	{\bf B7''} (dot-dashed) of Tables~\ref{scenario.B7} and 
	\ref{scenario.B7.Higgs} with $m_A=350$~GeV, $400$~GeV and $500$~GeV, 
	respectively, and longitudinal beam polarizations 
	$\mathcal{P}_{+} =\mathcal{P}_{-} = -0.3$.
}
\label{fig:asymm.prod.C}
\end{figure}

In  Fig.~\ref{fig:asymm.prod.C}a we compare the 
charge asymmetries $\mathcal{A}_{e}^{C}$~(\ref{achargeA})
for scenarios~{\bf B7}, {\bf B7'} and {\bf B7''},
that differ only in $m_A=\{350,400,500\}$~GeV,
as a function of $\sqrt{s}-m_A$.
We show the corresponding cross sections for 
$\mu^+\mu^-\to\charginoplus_1\charginominus_1$ in Fig.~\ref{fig:asymm.prod.C}b. 
For increasing Higgs masses their widths increase, 
and thus the interference of the $H$ and $A$ exchange amplitudes.
However, the maxima of the asymmetries are reduced by larger 
continuum contributions to the cross section.
For smaller Higgs masses, here $m_A=350$~GeV, the threshold effects 
are stronger.
Since a Dirac fermion-antifermion pair has negative intrinsic parity, 
and thus the $CP$ even $H$ resonance is p-wave suppressed, 
the peak cross section is found at $\sqrt{s}\approx m_A$,
where the asymmetry nearly vanishes. The asymmetry changes sign between 
the two resonances, whose mass difference is larger than their widths, 
due to the complex phases of the propagators. 
Its maximum is found at center of mass energies slightly above $m_H$ 
where the phases of the propagators
are roughly equal and the amplitudes of similar magnitude.
In Fig.~\ref{fig:asymm.prod.C}c we show the statistical significance 
$\mathcal{S}^C_e$
for an integrated effective luminosity 
$\mathcal{L}_{ef\!f} = 1~\rm{fb}^{-1}$.
We find statistical significances of $\mathcal{S}^C_e > 3$,
albeit not in the entire resonance region
for scenarios  {\bf{B7}} and {\bf{B7'}} with smaller  $m_A$. 

The asymmetries are also sensitive to a variation of $\tan\beta$. 
In the Higgs sector, increasing $\tan\beta$ results in 
larger $H$ and $A$ widths and smaller mass differences between $H$ and $A$.
This leads to a larger overlap of the two resonances, 
and thus to larger asymmetries $\mathcal{A}_{e}^{C}$ in the 
resonance region. 
In addition, since the couplings of the muons to the 
Higgs bosons~(\ref{eq:cHmu}) and (\ref{eq:cAmu}) are proportional to 
$\tan\beta$ in the Higgs decoupling limit~\cite{decoupling},
larger values of $\tan\beta$ imply smaller relative continuum 
contributions that enhance the asymmetries. 
On the contrary, for small $\tan\beta \lsim 5$ and $m_{H,A}< 2 m_t$, 
with $m_t$ the top quark mass,
the resonances practically do not overlap, see 
e.g.~\cite{Fraas:2003cx}, and the asymmetries cannot be measured. 
For $m_{H,A}> 2 m_t$, the resulting large $H$ and $A$ widths
may lead to an overlap of the resonances.
However, the combined effect of smaller Higgs-muon couplings
and the suppression of the cross section due to the large widths
imply a small resonant contribution with respect to the continuum 
and consequently only small asymmetries and statistical significances
are obtained.

\subsubsection{Chargino decay into a $W$ boson}

\begin{figure}[t]
\centering
\begin{picture}(14,8.2)
\put(0.2,1.){\includegraphics{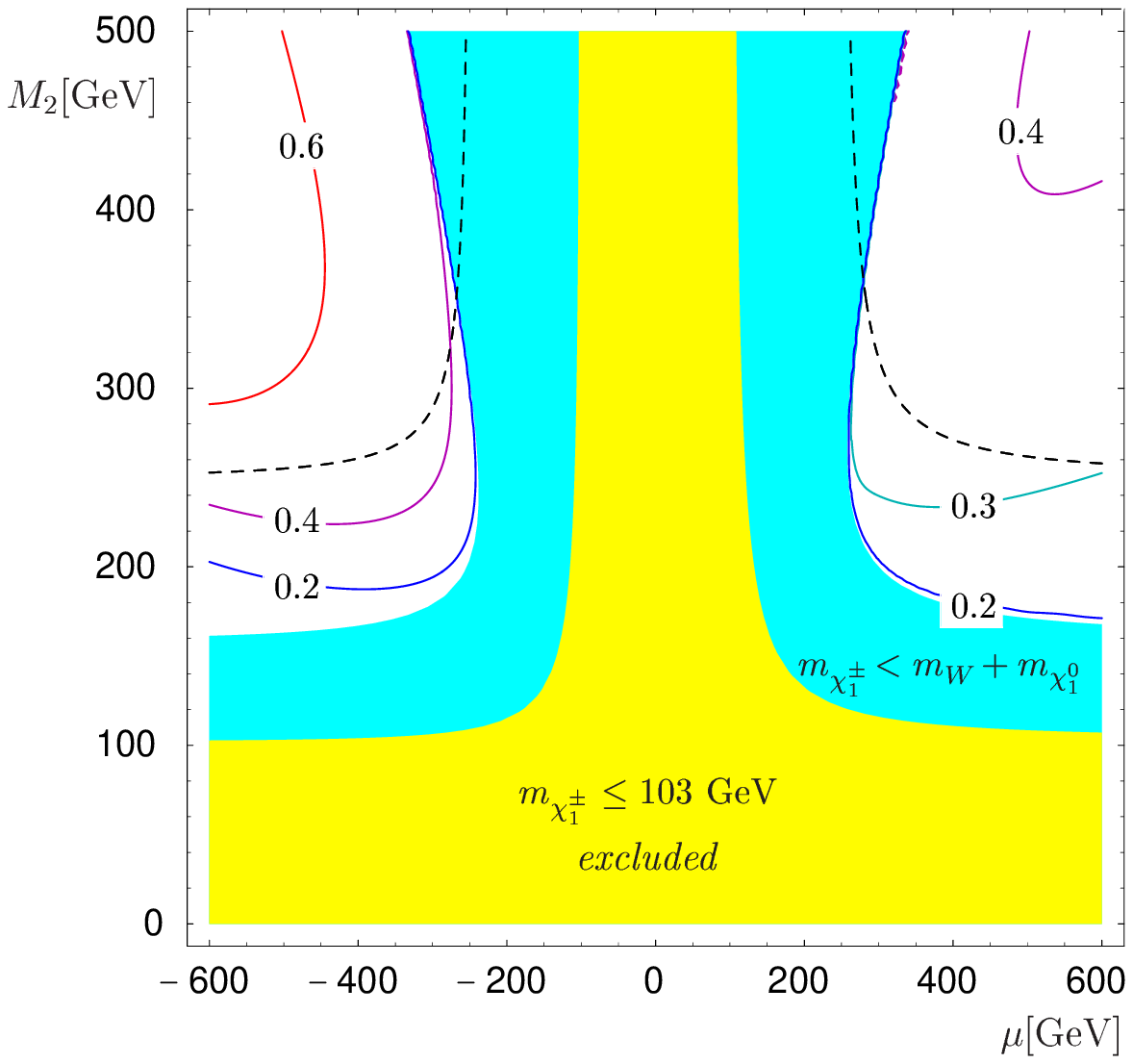}}
\end{picture}
\caption{\small 
	Contour lines of $\eta_{W^-}$~(\ref{etaW}) for 
	the decay $\charginominus_1\to W^-\neutralino_1$ 
	in the $\mu$--$M_2$ plane
	for $\tan\beta=10$. 
The dashed line indicates the kinematical limit
for $2m_{\chi_1^\pm}=\sqrt{s}=500$~GeV.
	The dark shaded area is kinematically forbidden by
	$m_{\chi^\pm} < m_{W}+m_{\chi_1^0}$.
	The light shaded area is experimentally excluded 
	by $m_{\chi_1^\pm}<103$~GeV.
}
\label{fig:etaw}
\end{figure}

If the sleptons are heavier than the charginos, the chargino decay
into a $W$ boson, $\tilde\chi_1^\pm \to W^\pm  \tilde\chi_1^0$,
might be the only allowed two-body decay channel.
In this case only the asymmetries of the energy
distribution of the $W$ boson, $\mathcal{A}_{W}^{C}$~(\ref{achargeA})
and $\mathcal{A}_{W^\pm}^{pol}$~(\ref{apolA}) are accessible. 
These asymmetries are reduced by a factor
$\eta_{W^\pm}$~(\ref{etaW}) 
with respect to
the asymmetries for leptonic chargino
decay modes. In Fig.~\ref{fig:contours}c we have shown the contour lines 
of the leptonic charge asymmetry  $\mathcal{A}_{e}^C$~(\ref{achargeA}) in
the $\mu$--$M_2$ plane for $\tan\beta=10$. The values of $\mathcal{A}_{e}^C$ have to be
multiplied by $\eta_{W^+}=-\eta_{W^-}$, which we show in
Fig.~\ref{fig:etaw}, to obtain the asymmetry  
$\mathcal{A}_{W}^C=\eta_{W^+}\times \mathcal{A}_{e}^C$.
Although the asymmetries are suppressed by $|\eta_{W^\pm}|\approx 0.2-0.4$,
and uncertainties in the energy measurement of the $W$ boson
lead to lower effective integrated luminosities,
statistics will be gained from large branching ratios, 
${\rm BR}(\tilde\chi_1^\pm \to W^\pm  \tilde\chi_1^0)=1$.

\subsection{Production of  $\chpm_1\chmp_2$}
\label{num:Production.of.chi1chi2}

\begin{table}
\renewcommand{\arraystretch}{1.2}
\caption{ Scenarios {\bf P1} and {\bf P2} for
	$\mu^+\mu^-\to\chpm_1\chmp_2.$
}
\begin{center}
	\begin{tabular}{|r|cc||r|cc|}
\hline
& {\bf P1} & {\bf P2} & & {\bf P1} & {\bf P2} 
\\
\hline
\hline
$\tan\beta$ 	& \hspace{.2cm}$10$\hspace{.2cm}		&\hspace{.2cm} $10$\hspace{.2cm}		& 	$m_{\chi^\pm_1}[\rm{GeV}]$ 	
&\hspace{.2cm} $138$\hspace{.2cm}	&\hspace{.2cm}$106$\hspace{.2cm}\\
\hline
$\mu[\rm{GeV}]$	& $-250$	& $-110$	&	$m_{\chi^\pm_2}[\rm{GeV}]$ 	& $281$	& $322$ 
\\
\hline
$M_2[\rm{GeV}]$	&  $150$	&  $300$	&	$m_{\chi^0_1}[\rm{GeV}]$	& $74$	& $89$
\\
\hline
$m_0[\rm{GeV}]$	&  $200$	&  $200$	&	$m_{\tilde{\nu}_\mu}[\rm{GeV}]$	& $232$	& $327$
\\
\hline
\hline
$m_A[\rm{GeV}]$ & $500$ 	& $500$ 	&  	 $\Gamma_A[\rm{GeV}] $		& $3.7$ 	& $3.4$
\\
\hline
$m_H[\rm{GeV}]$ &  $500.3$	& $500.4$	&	 $\Gamma_H[\rm{GeV}] $		&  $3.6$	& $3.3$
\\
\hline
\end{tabular}
\end{center}
\renewcommand{\arraystretch}{1.0}
\label{scenariosP12}
\end{table}
\begin{figure}[t]
\centering
\begin{picture}(16,6.7)
\put(10.4,5.9){$ \sigma(\mu^+\mu^-\to\chpm_1\chmp_2)$}
\put(-1.8,-10.6){\includegraphics{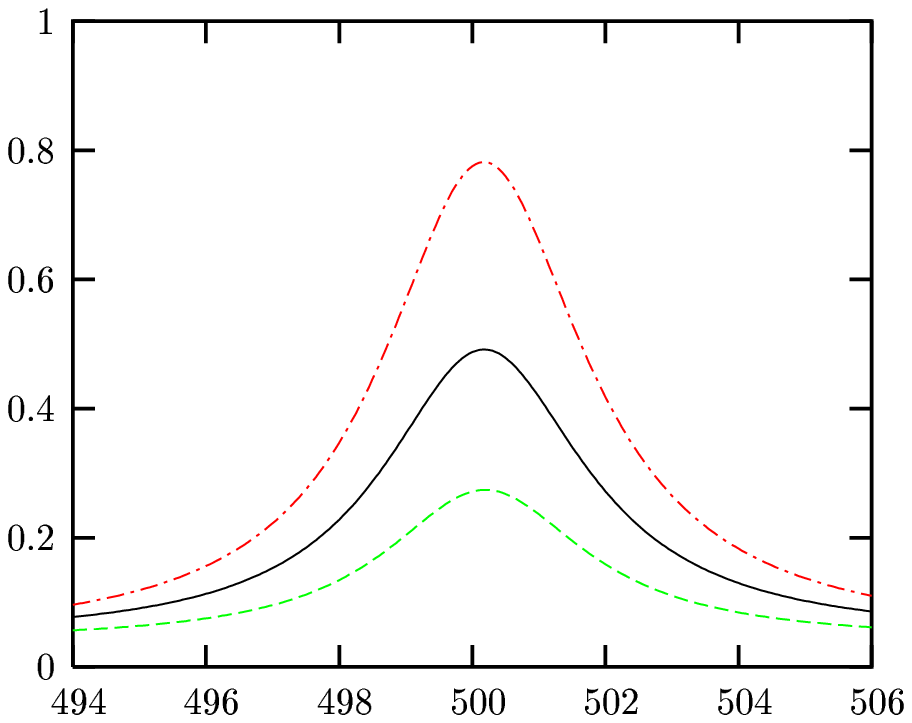}}
\put(1.09,-0.){ (a)}
\put(0.1,5){$ \scriptstyle\sigma[\rm{pb}]$} 
\put(5.8,.15){$ \scriptstyle \sqrt{s}[\rm{GeV}]$}
\put(2.5,5.9){$ \sigma(\mu^+\mu^-\to\chpm_1\chmp_2)$}
\put(6.0,-10.6){\includegraphics{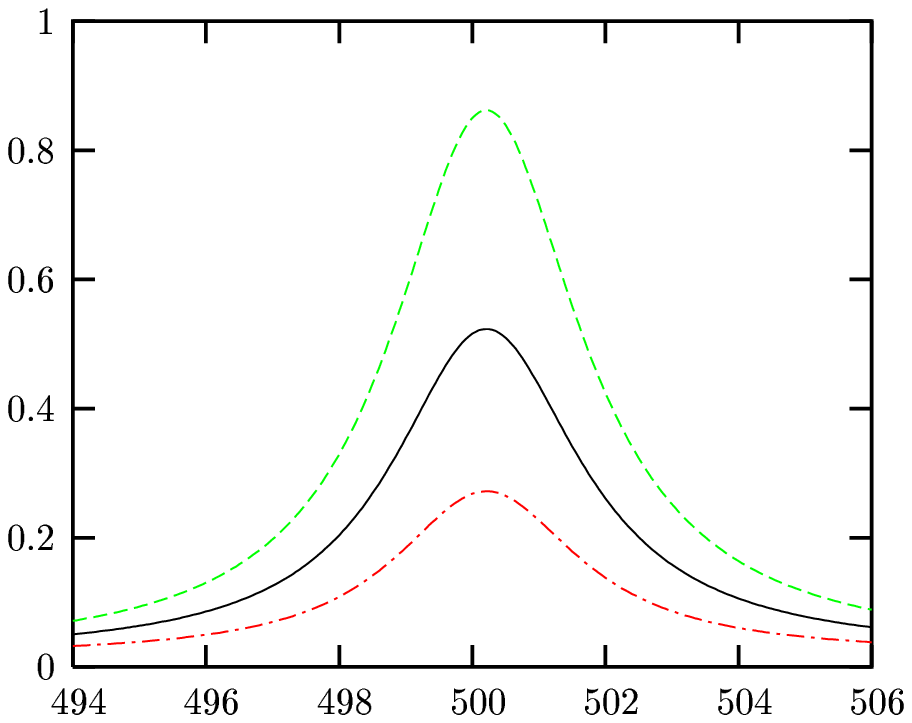}}
\put(8.97,0.){ (b)}
\put(7.99,5){$ \scriptstyle\sigma[\rm{pb}]$} 
\put(13.6,.15){$ \scriptstyle \sqrt{s}[\rm{GeV}]$}
\end{picture}
\caption{\small 
	$\mu^+\mu^-\to\chpm_1\chmp_2.$ Cross sections
	$ \sigma( \mu^+\mu^-\to\charginoplus_1\charginominus_2)$ (dashed) and 
	$ \sigma( \mu^+\mu^-\to\charginominus_1\charginoplus_2)$ (dash-dotted) 
	for longitudinal beam polarizations 
	$\mathcal{P}_+ =\mathcal{P}_- =\mathcal{P} = -0.3$, 
	and $ \sigma(\mu^+\mu^-\to\chpm_1\chmp_2)$ (solid) for $\mathcal{P} = 0$, 
	for scenario {\bf P1} {\bf (a)} and  
	scenario {\bf P2} {\bf (b)}, given in Table~\ref{scenariosP12}.
}
\label{fig:asymm.prod.DE}
\end{figure}
%

In Fig.~\ref{fig:asymm.prod.DE}a we show the cross sections for
$ \mu^+\mu^-\to\charginoplus_1\charginominus_2$  
production and for the charge conjugated process
$ \mu^+\mu^-\to\charginominus_1\charginoplus_2$
for scenario~{\bf P1}, given in Table~\ref{scenariosP12}.
The two cross sections are equal for unpolarized beams
and differ for polarized beams 
$\mathcal{P}_+ = \mathcal{P}_- = -0.3$.
In this case the  $H$--$A$ interference (\ref{pcpconsHA}) enhances the 
$\charginominus_1\charginoplus_2$ cross section and suppresses 
that for the conjugated process. 
The corresponding asymmetry $\mathcal{A}^C_{pr\!od}$~(\ref{aprodch1ch2})
of the two cross sections 
is $\mathcal{A}^C_{pr\!od}=-48\%$ at $\sqrt{s}=500$~GeV.
The asymmetry almost reaches its maximum absolute value of 
$|\mathcal{P}_{+} +\mathcal{P}_{-}|/(1+\mathcal{P}_{+}\mathcal{P}_{-})\approx 55\%$, 
here for $\mathcal{P}_+ =\mathcal{P}_- = -0.3$,
which would be obtained in the ideal case of vanishing continuum
contributions.
For scenario~{\bf P2}, shown in Fig.~\ref{fig:asymm.prod.DE}b, 
the $\charginominus_1\charginoplus_2$ production is instead suppressed 
by the $H$--$A$ interference and the $\charginoplus_1\charginominus_2$
production is enhanced, such that $\mathcal{A}^C_{pr\!od}=45\%$ changes sign.
In scenario~{\bf P1} ({\bf P2}) the lightest chargino has mainly 
gaugino (higgsino) character, i.e., the gaugino (higgsino) components are larger.
Since Higgs bosons couple to a gaugino-higgsino pair, 
the corresponding couplings (\ref{eq:cH}-\ref{eq:cLR}) 
transform as $c^{(\phi)}_{L,R\,ij}\leftrightarrow c^{(\phi)}_{L,R\,ji}$ 
under 
$M_2 \leftrightarrow |\mu|$.
This transformation relates the resonant amplitudes 
of $\charginoplus_1\charginominus_2$ and $\charginominus_1\charginoplus_2$ 
production for scenarios~{\bf P1} and {\bf P2},
which explains the different signs of
$\mathcal{A}^C_{pr\!od}$. 
Consequently for $M_2 = |\mu|$ the asymmetries vanish,
even for polarized beams.

\section{Summary and conclusions
  \label{Summary and conclusions}}
In the $CP$ conserving MSSM we have studied the s-channel interference 
of the $CP$ even and $CP$ odd neutral Higgs bosons $H$ and $A$ in chargino
production $\mu^+\mu^-\to\chmp_i\chpm_j$ with longitudinally polarized
beams. We have shown that the interference of $H$ and $A$ can be analyzed
for $\charginoplus_1\charginominus_1$ production using asymmetries in the energy distribution 
of the lepton or $W$ boson from the 
decay $\chpm_1\to\ell^\pm\sneutrino_\ell$, $\ell=e,\mu,\tau$,
or $\chpm_1\to W^\pm\neutralino_1$, respectively.
The asymmetries of the energy distributions are correlated to the 
longitudinal chargino polarizations.
For the production of two different charginos,
the $H$--$A$ interference can be analyzed using asymmetries of 
the $\charginoplus_1\charginominus_2$ cross section and its charge conjugate.
The asymmetries depend on the muon beam polarizations and 
thus vanish for unpolarized beams.
Since the asymmetries probe the $H$--$A$ interference, their 
measurement allows a determination of chargino couplings to the $H$
and $A$ bosons as well as a determination of the relative phase of
the couplings.
In a numerical study we have analyzed the production
of $\charginoplus_1\charginominus_1$ and $\chpm_1\chmp_2$ for different MSSM scenarios
and found asymmetries which are maximal for
nearly degenerate $H$ and $A$  bosons.
In the numerical analysis of the chargino cross sections and
branching ratios, we have shown that the asymmetries are accessible  
at a future muon collider with polarized beams. 
\clearpage

\begin{appendix}
\noindent{\Large\bf Appendix}
\setcounter{equation}{0}
\renewcommand{\thesubsection}{\Alph{section}.\arabic{subsection}}
\renewcommand{\theequation}{\Alph{section}.\arabic{equation}}

\setcounter{equation}{0}
\section{Chargino polarization vectors}
\label{spinvectors}

We choose a coordinate frame in the center of mass system (CMS)
such that the momentum of the chargino $\tilde\chi^{\pm}_j$ 
is given by
\begin{equation}
	p_{\chi^{\pm}_j}^{\mu}=(E_{\chi^{\pm}_j};0,0,|\vec{p}_{\chi^{\pm}_j}|),
\end{equation}
with
\begin{equation}
	E_{\chi_j^{\pm}} =\frac{s+m_{\chi^{\pm}_j}^2-m_{\chi^{\pm}_i}^2}{2 \sqrt{s}},\quad
    |\vec{p}_{\chi^{\pm}_j}|   =\frac{\lambda^{\frac{1}{2}}
		 (s,m_{\chi^{\pm}_i}^2,m_{\chi^{\pm}_j}^2)}{2 \sqrt{s}}. 
\end{equation}
The spin vectors 
of the chargino 
are then defined by 
\begin{equation}
	s_{\chi^{\pm}_j}^{1,\mu}= (0;1,0,0), \qquad	s_{\chi^{\pm}_j}^{2,\mu}= (0;0,1,0), 
	\qquad	s_{\chi^{\pm}_j}^{3,\mu}= 
	\frac{1}{m_{\chi^{\pm}_j}}(|\vec{p}_{\chi^{\pm}_j}|;0,0,E_{\chi^{\pm}_j}).
	\label{spinvec}
\end{equation}

\section{Chargino decay into $\tau $ and $W$ boson}
\label{decdens}

The interaction Lagrangians for chargino decay into a $\tau $, 
$\tilde\chi_j^{\pm} \to \tau^{\pm} ~ \tilde\nu_\tau^{(\ast)}$, and $W$
boson, $\tilde\chi_j^{\pm} \to W^{\pm}~\tilde\chi_k^0$, are,
respectively~\cite{HK}
\begin{eqnarray}
	{\cal L}_{\tau \sneutrino_\tau \tilde\chi^+} & = & 
		- g \bar{\tau}( V_{j1}  P_R -Y_{\tau}U_{j2}^*
			P_L)\tilde{\chi}_j^{+C} \sneutrino_\tau + \mbox{h.c.},\\
	{\cal L}_{W^-\tilde\chi^+\neutralino} & = & 
		g W^-_\mu  \bar{\tilde{\chi}}_k^{0} \gamma^\mu
			(O_{kj}^L P_L+O_{kj}^R P_R) 
		\tilde{\chi}_j^{+} \sneutrino_\ell
	+	\mbox{h.c.},
\end{eqnarray}
with the couplings 
\begin{eqnarray}
	O_{kj}^L	&=&	-\frac{1}{\sqrt{2}}	N_{k4}V_{j2}^\ast
				+			(\sin\theta_W N_{k1}+\cos\theta_W N_{k2})	V_{j1}^\ast,
\\
	O_{kj}^R	&=&	+\frac{1}{\sqrt{2}}	N_{k3}^\ast U_{j2}
				+			(\sin\theta_W N_{k1}^\ast+\cos\theta_W
					N_{k2}^\ast)	U_{j1},
\end{eqnarray}
and $Y_{\tau}= m_{\tau}/(\sqrt{2}m_W\cos\beta)$.
The $4\times 4$ unitary matrix $N$ diagonalizes the neutralino mass
matrix $Y$ in the basis 
$\{ \tilde{\gamma},\tilde{Z},\tilde{h}_1,\tilde{h}_2 \}$ with 
$N_{il}^\ast Y_{lm}N_{mj}^\dagger = \delta_{ij} m_{\chi^0_j}$~\cite{HK}.

The expansion coefficients of the chargino decay matrix~(\ref{rhoD}) 
for $\tilde\chi_j^+ \to \tau^+ ~ \tilde\nu_\tau$ are 
   \begin{eqnarray}
		D & = & \frac{g^2}{2} (|V_{j1}|^2 +Y_{\tau}^2|U_{j2}|^2)
		(m_{\chi_j^{\pm}}^2 -m_{\tilde{\nu}_\tau}^2 ),
   \label{DRtau} \\
      \Sigma^a_{D} &=&    
		 -g^2  (|V_{j1}|^2 -Y_{\tau}^2|U_{j2}|^2) 
		m_{\chi_j^{\pm}}(s^a_{\chi_j^{\pm}} \cdot p_{\tau}), 
		   \label{SigmaDtau}
\end{eqnarray}
and those for 
$\tilde\chi_j^+ \to W^+~\tilde\chi_k^0$ are  
   \begin{eqnarray}
      D & = & \frac{g^2}{2}(|O^L_{kj}|^2+|O^R_{kj}|^2)
	\left[
		m_{\chi_j^{\pm}}^2+m_{\chi_k^0}^2-2m_W^2+\frac{(m_{\chi_j^{\pm}}^2-m_{\chi_k^0}^2)^2}{m_W^2}
	\right]
\nonumber\\ & &
	-6 g^2 \mbox{Re}(O^L_{kj} O^{R\ast}_{kj})  m_{\chi_j^{\pm}} m_{\chi_k^0},
\label{DW}
\\[1mm]
      \Sigma^a_{D} &=& -g^2 (|O^L_{kj}|^2-|O^R_{kj}|^2)
\,
	\frac{	(m_{\chi_j^{\pm}}^2-m_{\chi_k^0}^2- 2 m_W^2)}{m_W^2}  
	\, m_{\chi_j^{\pm}} (s^a_{\chi_j^{\pm}}\cdot p_W).	
\label{SigmaDW}
   \end{eqnarray}
The coefficients $\Sigma^a_{D}$ for the charge conjugated processes, 
$\tilde\chi_j^- \to \tau^- ~ \tilde\nu_\tau^\ast$ and
$\tilde\chi_j^- \to W^-~\tilde\chi_k^0$,
are obtained by inverting the signs of~(\ref{SigmaDtau})
and~(\ref{SigmaDW}), respectively. 

For the chargino decay $\tilde\chi_j^{\pm} \to W^{\pm}~\tilde\chi_k^0$ 
the energy limits of the $W$ boson are 
$E_W^{max(min)}= \bar{E}_W \pm \Delta_W$, see~(\ref{kinlimits}),
with
\begin{eqnarray}
	\bar{E}_W &=& \frac{E_W^{max}+E_W^{min}}{2} = 
	\frac{ m_{\chi^{\pm}_j}^2+m_W^2-m_{\chi_k^0}^2}{2 m_{\chi^{\pm}_j}^2} E_{\chi^{\pm}_j},
		\label{ehalfW}
\\
	\Delta_W &=& \frac{E_W^{max}-E_W^{min}}{2} = \frac{ \sqrt{
	\lambda(m_{\chi^{\pm}_j}^2,m_W^2,m_{\chi_k^0}^2)
	}}{2 m_{\chi^{\pm}_j}^2} |\vec{p}_{\chi^{\pm}_j}|.
\label{edifW}
\end{eqnarray}
The factor $\eta_{\lambda^{\pm}}$~(\ref{etal})
for the decay 
$\tilde\chi_j^{\pm} \to \tau^{\pm} ~ \tilde\nu_\tau^{(\ast)}$
is given by
\begin{eqnarray}
	\eta_{\tau^{\pm}} = \pm\frac{ |V_{j1}|^2 -Y_{\tau}^2|U_{j2}|^2}
	{ |V_{j1}|^2 +Y_{\tau}^2|U_{j2}|^2}.
\label{etatau}
\end{eqnarray}
For the  decay
$\tilde\chi_j^{\pm} \to W^{\pm}~\tilde\chi_k^0$
we find
\begin{eqnarray}
	\eta_{W^{\pm}}&=&   \pm   \frac{(|O^L_{kj}|^2-|O^R_{kj}|^2) f_1}
			 	{(|O^L_{kj}|^2+|O^R_{kj}|^2) f_2  
					+ \mbox{Re}\{O^L_{kj} O^{R\ast}_{kj}\}  f_3},
\label{etaW}
\end{eqnarray}
with
\begin{eqnarray}
	f_1&=&	(m_{\chi_j^{\pm}}^2-m_{\chi_k^0}^2- 2 m_W^2) \sqrt{
	\lambda(m_{\chi^{\pm}_j}^2,m_W^2,m_{\chi^0_k}^2)
	}
, \nonumber\\
f_2&=&	(m_{\chi_j^{\pm}}^2+m_{\chi_k^0}^2- 2 m_W^2)~ m_W^2 + 
(m_{\chi_j^{\pm}}^2-m_{\chi_k^0}^2)^2, \nonumber\\
	f_3&=&	-12~ m_{\chi_j^{\pm}}~ m_{\chi_k^0}~ m_W^2.\nonumber
\end{eqnarray}
The coefficients $\eta_{\tau^{\pm}}$ and $\eta_{W^{\pm}}$
depend on the $\tau$ and $W$ couplings to the charginos, as well as on
the chargino and neutralino masses, which could be measured at the 
international linear collider (ILC) with 
high precision~\cite{TDR,Nojiri.Boos.Martyn}.

\end{appendix}



\end{document}